\documentclass[twocolumn]{aastex631}

\usepackage{enumitem}

\newcommand{\tdisc }{t_{\rm disc}}

\newcommand{\Mearth}{M_\oplus}

\newcommand{\yr}{{\rm yr}}

\definecolor{darkpurple}{HTML}{561977}

\shorttitle{Migration traps \& the Kepler dichotomy}
\shortauthors{Zawadzki et al.}

\begin{document}

\title{Migration traps as the root cause of the Kepler dichotomy}

\author{Brianna Zawadzki}
\affiliation{Department of Astronomy and Astrophysics, 525 Davey Lab, The Pennsylvania State University, University Park, PA 16802, USA}
\affiliation{Center for Exoplanets \& Habitable Worlds, 525 Davey Lab, The Pennsylvania State University, University Park, PA 16802, USA}

\author{Daniel Carrera}
\affiliation{Department of Physics and Astronomy, Iowa State University, Ames, IA, 50010, USA}
\affiliation{Department of Astronomy and Astrophysics, 525 Davey Lab, The Pennsylvania State University, University Park, PA 16802, USA}
\affiliation{Center for Exoplanets \& Habitable Worlds, 525 Davey Lab, The Pennsylvania State University, University Park, PA 16802, USA}

\author{Eric B. Ford}
\affiliation{Department of Astronomy and Astrophysics, 525 Davey Lab, The Pennsylvania State University, University Park, PA 16802, USA}
\affiliation{Center for Exoplanets \& Habitable Worlds, 525 Davey Lab, The Pennsylvania State University, University Park, PA 16802, USA}
\affiliation{Institute for Computational \& Data Sciences, The Pennsylvania State University, University Park, PA 16802, USA}
\affiliation{Institute for Advanced Study, 1 Einstein Drive, Princeton, NJ 08540, USA}
\affiliation{Center for Computational Astrophysics, Flatiron Institute, New York, NY 10010, USA}



\begin{abstract}

It is often assumed that the ``Kepler dichotomy'' --- the apparent excess of planetary systems with a single detected transiting planet in the Kepler catalog --- reflects an intrinsic bimodality in the mutual inclinations of planetary orbits. After conducting 600 simulations of planet formation followed by simulated Kepler observations, we instead propose that the apparent dichotomy reflects a divergence in the amount of migration and the separation of planetary semimajor axes into distinct ``clusters''. We find that our simulated high-mass systems migrate rapidly, bringing more planets into orbital periods of less than 200 days. The outer planets are often caught in a migration trap --- a range of planet masses and locations in which a dominant co-rotation torque prevents inward migration --- which splits the system into two clusters. If clusters are sufficiently separated, the inner cluster remains dynamically cold, leading to low mutual inclinations and a higher probability of detecting multiple transiting planets. Conversely, our simulated low-mass systems typically bring fewer planets inside 200 days, forming a single cluster that quickly becomes dynamically unstable, leading to collisions and high mutual inclinations. We propose an alternative explanation for the apparent Kepler dichotomy in which migration traps during formation lead to fewer planets inside the Kepler detection window, and where mutual inclinations play only a secondary role. If our scenario is correct, then Kepler's STIPs (Systems with Tightly-packed Inner Planets) are a sample of planets that escaped capture by co-rotation traps, and their sizes may be a valuable probe into the structure of protoplanetary discs.

\end{abstract}



\section{Introduction} \label{sec:intro}

NASA's Kepler mission has revolutionised exoplanet science with the discovery of thousands of super-Earth size planets in close-in orbits around their host stars \citep{Borucki_2010,Batalha_2013}. One of the most interesting puzzles arising from Kepler data is the apparent excess of single transit systems, known as the Kepler dichotomy, relative to what might be expected based on the number of systems with multiple transiting planets detected. Briefly, several models for the population of planetary systems that reproduces key properties of the multiple-transit systems predict fewer systems with a single detected transiting planet that observed \citep{Lissauer_2011,Johansen_2012,Fabrycky_2014,Ballard_2016,Mulders_2018,He_2019}.

One early model for the Kepler dichotomy was a literal excess in the number of systems with a single planet within the range of orbital periods and sizes detectable by Kepler.  \citet{He_2019} rejected this model by simultaneously modeling the number of planets detected, the transit observables that probe system architectures, and the Kepler detecting and vetting efficiency.  They showed that after setting aside the fraction of stars needed to explain the multiple transiting planet systems, there were not enough stars remaining to explain the Kepler dichotomy --- even if 100\% of the remaining stars were assigned a single planet.

Perhaps the most commonly proposed explanation for the Kepler dichotomy has been that the systems with a single detected planet --- the ``Kepler singles'' --- are a distinct population of dynamically hot multi-planet systems with high mutual inclinations such that very few viewing angles allow for multiple transits. For example, \citet{Moriarty_2016} have sought to reproduce the Kepler dichotomy by invoking a bimodal distribution of mutual inclinations so that planetary systems belong to either a dynamically hot or dynamically cold population. This could be caused by spin-orbit misalignment, where stellar rotation exerts a torque on misaligned planetary orbits, exciting mutual inclinations between planets and sometimes causing planets to become dynamically unstable \citep{Spalding_2016}.
Similarly, others have proposed that a bifurcation of formation pathways in which some planetary systems experience dynamical instabilities and some do not could also explain the Kepler dichotomy \citep{Johansen_2012, Izidoro_2017, Izidoro_2021}. This is consistent with evidence that single-transiting planets generally have higher eccentricities than observed multiplanet systems \citep{VanEylen_2019, He_2020}. Both proposals postulate that the observed dichotomy is caused by an inherent divergence in the formation history of planetary systems, leading to distinct differences in system architectures and especially mutual orbital inclinations.

It has also been proposed that the Kepler dichotomy is merely an artifact from the Kepler detection pipeline; \citet{Zink_2019} inferred this by using Kepler field stars with artificially injected planet signals, as the pipeline experiences a significant loss in transit detection efficiency after the first transit (with the strongest signal) has been detected. However, their model for the detection efficiency shows that variations with multiplicity are only significant for planets near the threshold of detection.  Since most planets detected by Kepler are well above the detection threshold, the \citet{Zink_2019} results may be due to a combination of their detection efficiency model and their model for the distribution of planetary systems.

Others have explored the idea that Kepler singles descended from systems with tightly-packed inner planets (STIPs) that destabilized over billions of years, consolidating or destroying the system and leaving few or no surviving planets \citep[e.g.][]{Pu_2015, Volk_2015}. While there is a rich literature of further possible explanations for the apparent Kepler dichotomy, such as a flat inner disk \citep{Bovaird_2017}, most literature on the Kepler dichotomy relies on the fact that planetary systems with high mutual inclinations are more likely to be identified as single-planet systems \citep[e.g.][]{He_2019,Moriarty_2016, Dong_2017, Mulders_2019, Mulders_2020, He_2020,Millholland_2021}. This work explores other factors that may contribute to the apparent Kepler dichotomy.


\subsection{Planet formation}
\label{sec:intro:formation}

In this study we generate a population of planetary systems using a modern planet formation model, so as to explore factors that may contribute to the apparent Kepler dichotomy. 
This section provides a succinct overview of key processes included in our model. Planet formation begins with small solid particles embedded in a gaseous protoplanetary disc around a young star. Gas turbulence leads to collisions between particles \citep{Weidenschilling_1977} which, depending on the collision speed and material strength, can result in coagulation or fragmentation of these dust grains \citep{Guettler_2010,Zsom_2010,Blum_2018}. 

The solids likely overcome the fragmentation barrier via a hydrodynamic process --- the \textit{streaming instability} --- which concentrates solids into dense filaments that undergo gravitational collapse \citep[e.g.][]{Youdin_2005,Johansen_2007,Bai_2010,Carrera_2015,Yang_2017,Nesvorny_2019}, resulting in the formation of 1-100 km bodies called planetesimals. Planetesimals then collide and grow, first through runaway growth \citep{Greenberg_1978,Wetherill_1989,Kokubo_1996}, followed by slower oligarchic growth \citep[e.g.][]{Kokubo_2000,Thommes_2003,Chambers_2006}. These processes may also be aided by the aerodynamically enhanced capture of pebbles from the disc \citep{Lambrechts_2012,Levison_2015}. Once a planet reaches roughly the size of Mars, it can alter the nearby structure of the disc: Lindblad resonances cause a spiral-like overdensity near the planet, trapping some gas in horseshoe orbits near the planet \citep{Goldreich_1979}. This structure induces a torque on the planet which causes the planet to experience \textit{Type I migration}. Typically the Lindblad torque is dominant and negative, leading to inward planet migration. It is expected that most observed close-in planets with significant gas envelopes experienced inward migration before arriving at their final orbital location \citep{Inamdar_2015}. However, for a roughly Earth-size planet the (positive) co-rotation torque can significantly slow down migration or even reverse its direction \citep{Papaloizou_2000,Paardekooper_2010,Paardekooper_2011}.

As planets migrate they experience dampening of orbital eccentricity and inclination \citep{Tanaka_2004}. Planets can also be captured into long chains of mutual mean motion resonances \citep[e.g.][]{Cresswell_2006,Cresswell_2008,Terquem_2007}. While a lone planet would stop migrating once it reaches the inner edge of the disc, a \textit{resonant chain} will continue to push the planet inward as long as the outer planets in the chain experience sufficiently strong torques. The result is that resonant chains can push multiple planets well past the inner edge of the protoplanetary disc \citep{Brasser_2018,Carrera_2019}.


\subsection{Dynamical stability}
\label{sec:intro:stability}

Previous studies have suggested that the Kepler dichotomy may be connected to dynamical stability.  We say that a planetary system is \textit{Hill stable} if planet orbits never cross. Gravitational interactions between closely-spaced planets often cause sudden changes in orbital elements and rapid chaotic evolution that inevitably leads to planet ejections and/or collisions \citep[e.g.][]{Chambers_1996,Rasio_1996,Chatterjee_2008,Juric_2008}. While the gas disc is still present, planets experience dampening of orbital eccentricities and inclinations \citep{Papaloizou_2000,Tanaka_2004}, which help stabilise the system. However, once the gas dissipates, the system will often evolve chaotically and become dynamically unstable, beginning a new phase of collisions \citep{Izidoro_2017}.

An intuitive way to think about dynamical stability is in terms of the total angular momentum deficit (AMD) of the system. Planets readily exchange angular momentum (and therefore AMD) through secular interactions. 
A system is said to be \textit{AMD unstable} if there is enough AMD in the system to allow two orbits to either cross or enter a region of mean motion resonance overlap \citep{Laskar_2017,Petit_2017}. Otherwise the system is said to be \textit{AMD stable}. AMD stability is a conservative criterion --- a system that is AMD stable is very likely to be Hill stable, but a system that is AMD unstable may be able to survive for several Gyr before experiencing orbit crossing. A good example is our own Solar System. The Solar System is AMD unstable because there is more than enough AMD in Jupiter's orbit to make the orbits of the terrestrial planets cross. One can take a more nuanced view of stability and note that a planetary system that is AMD unstable \textit{in principle} may survive for several Gyr if it is difficult to transfer AMD between planets. It is worth noting that some authors have shown that there is a small probability of Mercury one day colliding with the Sun or Venus \citep{Batygin_2008}. Later in this paper we will propose that when planet formation produces two distinct planet clusters with a large separation between them, the system is more likely to be stable than if all the planets form a single cluster. This can be understood in terms of the difficulty of transferring AMD from one planet cluster to another.


\subsection{The origin of system architectures}
\label{sec:intro:summary}

Planet migration and dynamical instabilities are the underlying processes that shape the architectures of planetary systems. Therefore, simplistic simulations (i.e., those that ignore migration, traps, instabilities, etc.) may not give accurate predictions for the key architectural properties of planetary systems. Such properties include planet semimajor axes, orbital separations, inclinations, radii, and potentially other properties that affect the apparent transit multiplicities. In this work we use a sophisticated models for both planet formation and the observation biases of the Kepler mission. We use these simulations to investigate the relative contributions of different effects on the relative abundance of apparently single and multiple planet systems.

Another key variable is the amount of solid mass available to form planets, both in terms of the total solid mass in the disk and the solid surface density in a given area of the disk. Other studies have shown that these factors can have a significant impact on the architecture of a final planetary system \citep[e.g]{Dawson_2016,Moriarty_2016,MacDonald_2020}. We use simulations that vary both the total solid mass in the disk and the solid surface density profile to investigate the impact of different possible disks on the Kepler dichotomy.

The paper is organised as follows. In section \ref{sec:methods} we summarise our planet formation simulations and how we model detection biases of the Kepler mission. We present our results in section \ref{sec:results}. The implications of those results are discussed in section \ref{sec:discussion}, and we conclude in section \ref{sec:conclusion}.

%
%
\section{Methods}
\label{sec:methods}

We use the same planet formation model as \citet{Carrera_2019}, but with different initial conditions. We use a modified version of the \textsc{mercury} N-body code's hybrid integrator \citep{Chambers_1999}, adding a user-defined force that implements the Type-I disc torques caused by a protoplanetary disc (Section \ref{sec:model:torques}). We model the disc as a 1-dimensional steady state accretion disc with alpha-viscosity $\nu = \alpha c_{\rm s} H$ \citep{Shakura_1973}, so that the accretion rate is given by
\begin{equation}
	\dot{M}_{\rm acc} = 3\pi\nu\Sigma =  3 \pi\alpha c_{\rm s} H \Sigma.\\
\end{equation}
Here, $c_{\rm s} = \sqrt{kT/\mu}$ is the isothermal sound speed, $k$ is the Boltzmann constant, $\mu$ is the mean molecular mass, $T$ is the disc temperature, $H = c_{\rm s}/\Omega$ is the scale height, $\Omega$ is the Keplerian orbital frequency, and $\Sigma$ is the gas surface density. All of our simulations have a stellar mass of $1 M_\odot$. We implement the disc temperature profile of \citet{Bitsch_2015}, following their prescription for the accretion rate
\begin{equation}\label{eqn:Mdot}
	\log_{\rm 10}\left( \frac{\dot{M}_{\rm acc}}{M_\odot / \yr} \right)
    = - 8 - 1.4\,\log_{\rm 10}\left( \frac{\tdisc + 10^5 \yr}{10^6 \yr} \right)
\end{equation}

\noindent with the nominal value $\alpha = 0.0054$. Together, these formulas fully define the disc structure (i.e. $T(r,t), \Sigma(r,t), H(r,t)$, etc.). All of our simulations begin when the disk is 1 Myr old. We allow the disc to evolve until $t_{\rm disc} = 5$ Myr, at which point we hold $T(r)$ fixed and lower the disc mass exponentially from 5 to 5.1 Myr with an e-folding timescale of $10^4$ years. At 5.1 Myr we remove the disc entirely and the run proceeds as a pure N-body simulation up to 100 Myr.

\subsection{Experiment setup}
\label{sec:methods:init}

We run three sets of 200 simulations. Each simulation begins with $135$, $155$, or $229$ embryos with masses of $0.05 - 0.20 M_\oplus$ and dynamical separations of at most 3 mutual Hill radii,
\begin{eqnarray}
    \Delta_j    &=& \frac{a_{j+1} - a_j}{R_{j,\rm H}} \le 3 \\
    R_{j,\rm H} &=& \left(\frac{2 m_j}{3 M_\star}\right)^{1/3} a_j,
\end{eqnarray}
where $m_j$ and $a_j$ are the mass and semimajor axis of the $j^{\rm th}$ embryo. The exact masses and separations of the embryos are determined by a target surface density profile,
\begin{equation}
    \Sigma_{\rm emb} = \Sigma_0 \left( \frac{r}{\rm AU} \right)^{- \gamma}.
\end{equation}

Each set of runs begins with a different choice of $\gamma$. As in \citet{Carrera_2019}, we assume that planetesimal formation occurs early and that planetesimals do not experience significant migration until they reach the size of planetary embryos, which is when our simulations begin. For that reason, in all our models $\Sigma_0$ is in the order of $\approx 0.01\Sigma_{\rm gas}(r = 1{\rm AU})$ at $t_{\rm disc} = 10^4$ yr. The innermost embryo is always placed at $a_1 = 1$ AU and given an initial mass of $m_1 = 0.05 M_\oplus$. The next few embryos are placed at $\Delta = 3$, and given a mass of
\begin{eqnarray}
    m_{j+1} &=& \int_{a_j}^{a_{j+1}} 2\pi r \Sigma_{\rm emb}(r)\;{\rm d}r\\
    a_{j+1} &=& a_j + 3 R_{j,\rm H}.
\end{eqnarray}
With each step the embryo mass increases slightly until it reaches $m_j = 0.2 M_\oplus$. At that point, we stop increasing the embryo mass and instead allow the embryos to be more closely spaced than $\Delta = 3$,

\begin{equation}
    m_{j+1} = 0.2 M_\oplus 
            = \int_{a_j}^{a_{j+1}} 2\pi r \Sigma_{\rm emb}(r)\;{\rm d}r,
\end{equation}
and solve for $a_{j+1}$. This setup ensures the initial embryos are both small and closely spaced, so that our simulations model the final formation of isolation-mass bodies. We stop iterating when the $j^{\rm th}$ embryo reaches 6 AU. Table \ref{tab:threemodels} shows the $\gamma$ values that we selected for each model, along with the resulting total embryo mass $M_{\rm{total}}$ and initial number of embryos $N_{\rm{emb}}$. We chose those parameters such that 
\begin{itemize}
    \item Models A and B have the same $\Sigma_0$,
    \item Models A and C have the same $\gamma$, and
    \item Models B and C have the same mass between $1-6$ AU.
\end{itemize}
\noindent This allowed us to simultaneously test the effects of $\Sigma_0$, $\gamma$, and total solid mass.

\begin{table}
  \centering
  \caption{We run three sets of 200 planet formation simulations. Each simulation starts with roughly Mars-sized embryos placed between 1 and 6 AU. The embryo masses and separations are arranged to follow a powerlaw surface density $\Sigma_{\rm{emb}} = \Sigma_{0} \; (r/{\rm{AU}})^{-\gamma}$. In this table, $\Sigma_{{\rm{gas}},0}$ is the gas surface density at 1 AU when $\tdisc = 10^{4}$ yr, $M_{\rm{total}}$ is the total mass in embryos, and $N_{\rm{emb}}$ is the number of embryos.}
  \label{tab:threemodels}
  \begin{tabular}{ccccc}
  Model & $100\;\Sigma_0/\Sigma_{{\rm gas},0}$ & $\gamma$ & $M_{\rm total}$ & $N_{\rm emb}$\\
  \hline
  A     & 1.000 & 0.5 & 43.8$\Mearth$ & 229\\
  B     & 1.000 & 1.0 & 24.1$\Mearth$ & 135\\
  C     & 0.563 & 0.5 & 24.8$\Mearth$ & 155\\
  \end{tabular}
\end{table}

We run 200 simulations of each model to obtain a statistically useful sample. In each iteration the embryos are given small but non-zero initial eccentricities ($e = 0.002$) and inclinations ($I = 0.10^\circ$). All other orbital elements ($\omega$, $\Omega$, $M$) are chosen uniformly randomly between $0^\circ$ and $360^\circ$.

Model A is based on the assumption that $\Sigma_{\rm emb} \approx Z\;\Sigma_{\rm gas}$, where $Z = 0.01$ is the disc metallicity, and that the initial planetesimals that gave rise to the embryos formed when the disc was very young. In the disc model of \citet{Bitsch_2015}, $\Sigma_{\rm gas} \propto r^{-0.5}$ near 1 AU, so we set $\gamma = 0.5$ as the powerlaw for model A. To obtain models B and C we consider the possibility that radial drift of solids may steepen the solid surface density profile \citep[e.g.][]{Birnstiel_2012}, or that planetesimal formation may be less efficient or occurs less early later than assumed in model A.


\begin{figure}
  \gridline{\fig{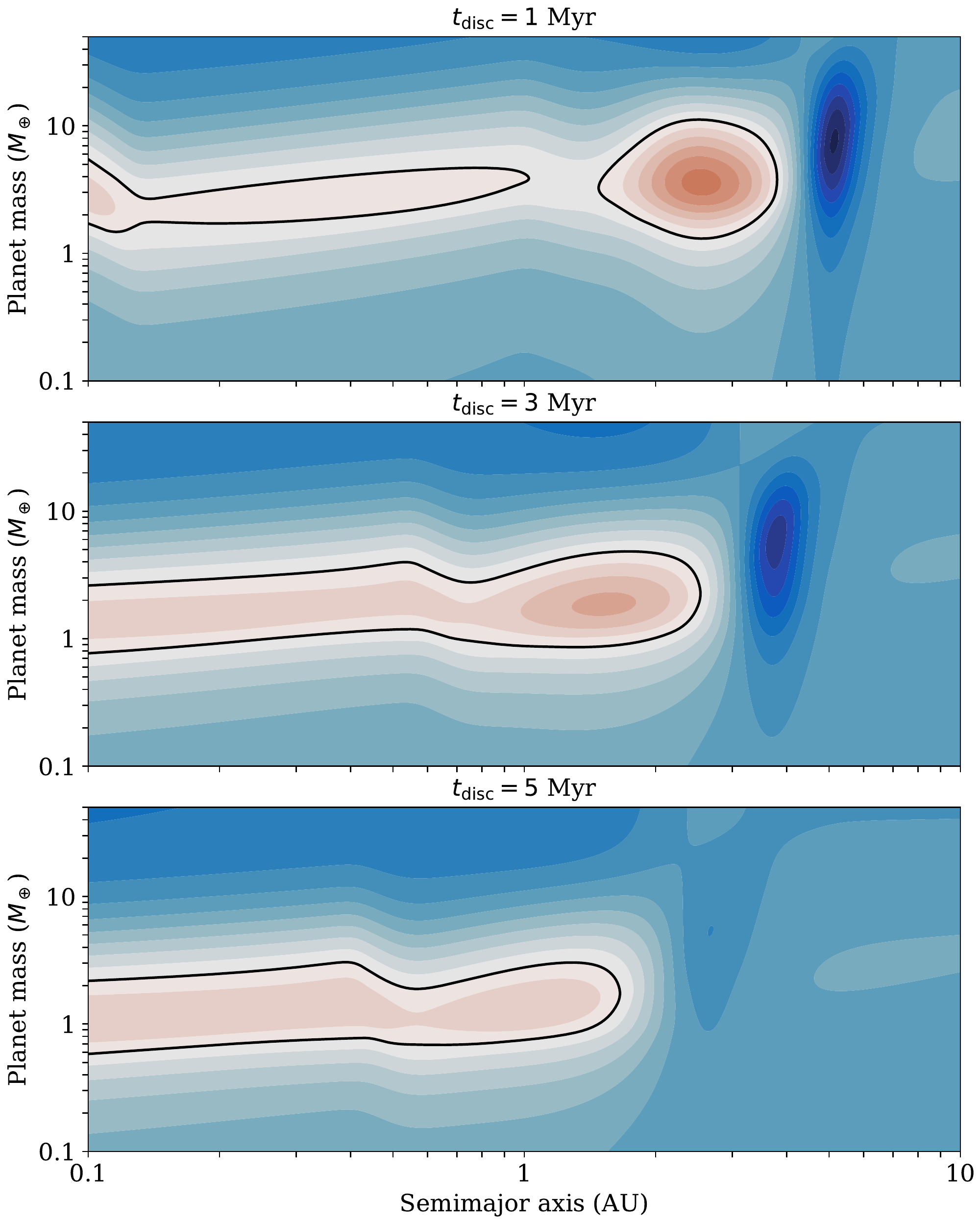}{0.49\textwidth}{}}
  \vspace{-0.3in}
  \gridline{\fig{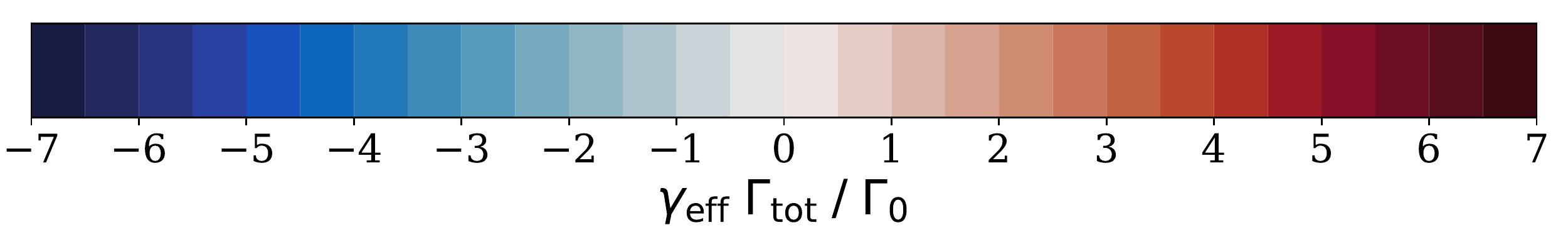}{0.43\textwidth}{}}
  \caption{Evolution of torques exerted on a planet by the disc from $t_{\rm{disc}}=1$ Myr to $t_{\rm{disc}}=5$ Myr. The colour scale shows the torque normalised by the effective adiabatic index $\gamma_{\rm eff}$ and the scaling factor $\Gamma_0 = (q/h)^2\,\Sigma\,r^4 \Omega^2$, where $q = m_{\rm pl}/M_\star$ and $h = H/r$. In most cases, the net torque is negative, driving inward migration. Inside the black boundaries, however, the net torque is positive and planets migrate outward. This occurs when the (positive) co-rotation torque, which depends strongly on the planet mass, overwhelms the (negative) Lindblad torque. Migration traps occur when a forming planet is at a mass and disk location that prevents inward migration.}
  \label{fig:torques}
\end{figure}

\subsection{Disc torques}
\label{sec:model:torques}

Our planet formation model only produces sub-Neptune mass planets, which are not massive enough to form gaps or significantly alter the structure of the disc. As a result, we only need to consider Type I migration in our model. In Type I migration, the direction and magnitude of the planet's migration is determined by a combination of the Lindblad and co-rotation torques,
\begin{equation}
    \Gamma_{\rm total} = \Gamma_{\rm L} + \Gamma_{\rm C},
\end{equation}
where $\Gamma_{\rm L}$ is the Lindblad torque, and $\Gamma_{\rm C}$ is the co-rotation torque. The Lindblad torque is caused by spiral density waves induced by Lindblad resonances; it is almost always negative (i.e. drives inward migration) and it is usually the dominant torque. The co-rotation torque is caused by gas in horseshoe orbits near the planet. As fluid elements execute the U-turns in the orbit they form a density gradient that give the planet a positive torque. For an introduction, see \citet{Kley_2012} and \citet{Armitage_2007}.

The formulas implemented in our code were derived by numerous authors including \citet{Papaloizou_2000,Tanaka_2004,Cresswell_2006,Cresswell_2008,Paardekooper_2010,Paardekooper_2011,Coleman_2014,Fendyke_2014}. Since the full set of formulas is long and some are fairly complex, we refer the reader to the appendix of \citet{Carrera_2019}, where they are described in full.

Here we simply note that the co-rotation torque has an interesting dependence on planet mass that can be seen in Figure \ref{fig:torques}. The figure shows how the disc torques vary with planet mass, semimajor axis, and disc age. The co-rotation torque is maximised when the libration period of the horseshoe is equal to the thermal diffusion timescale of the gas in the disc. As a result, there is a particular planet mass range where the co-rotation torque is strongest and exceeds the Lindblad torque to drive outward migration. For higher and lower masses, the co-rotation torque is smaller and the Lindblad torque generally dominates. Therefore, it is possible for some planets to become caught in a migration trap while others continue to migrate inward, which would give rise to distinct planet clusters.

\subsection{Observational biases}
\label{sec:methods:obs}

\begin{figure}
  \gridline{\fig{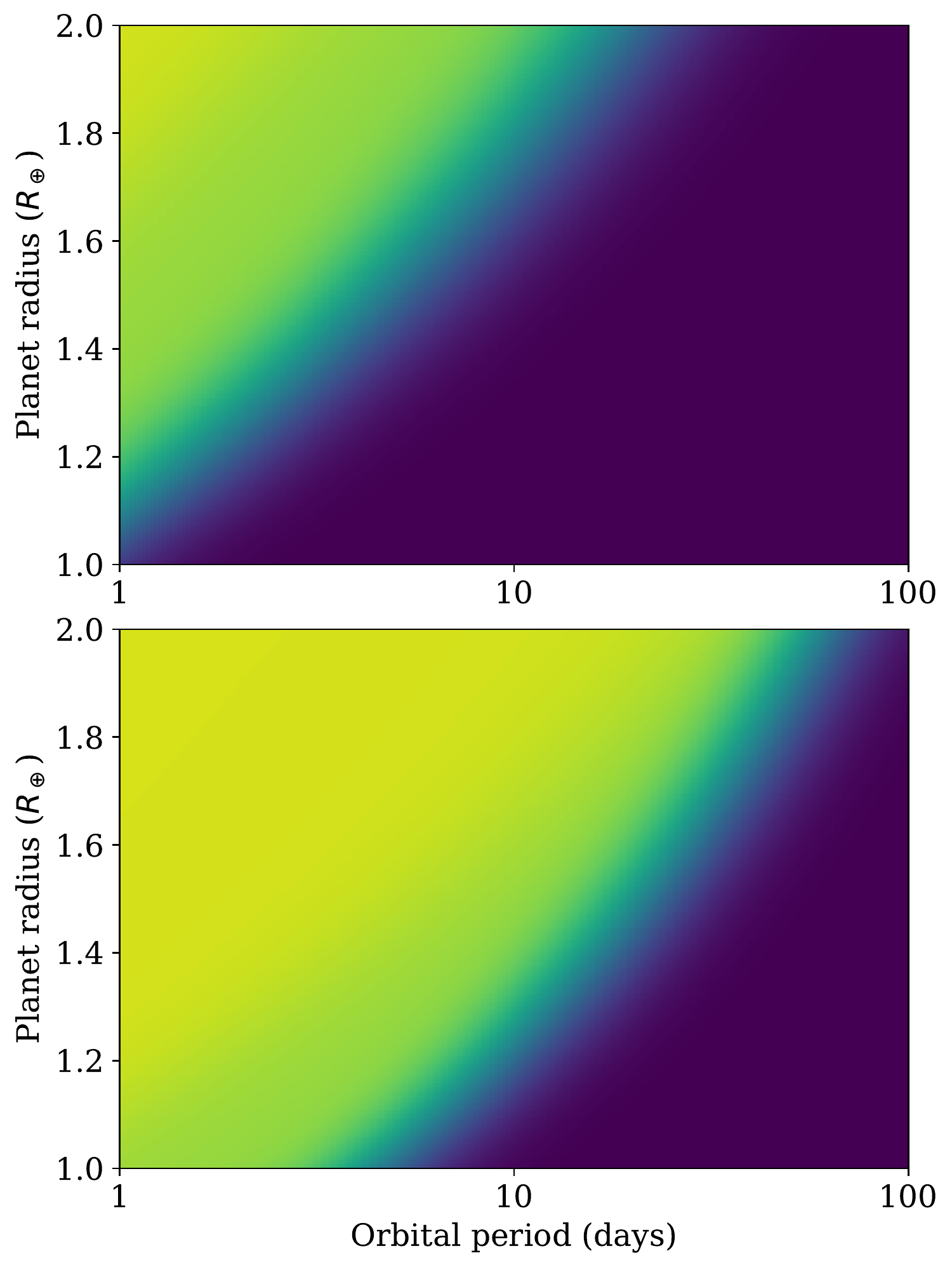}{0.47\textwidth}{}}
  \vspace{-0.3in}
  \gridline{\fig{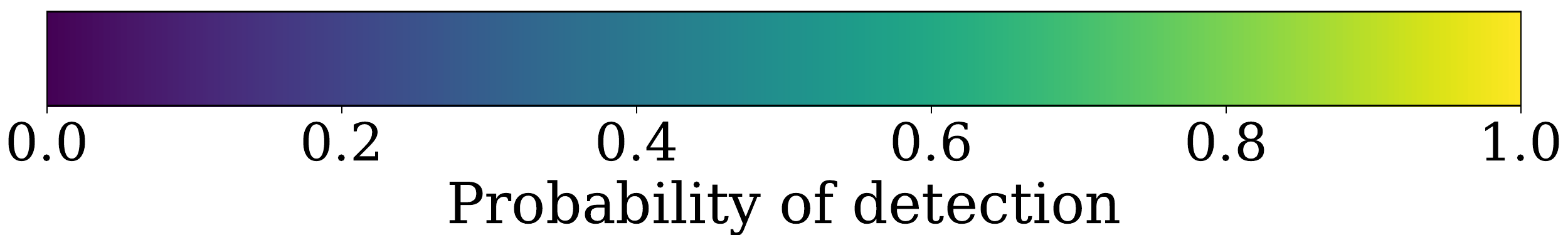}{0.435\textwidth}{}}
  \caption{The probability of detecting a transiting planet around two sample G dwarf stars for different planet radii and orbital periods. The detection probability depends on the transit depth, period, and stellar noise. In this example the two stars have comparable stellar noise, but the star in the top panel has a larger radius, which reduces the transit depth.}
  \label{fig:det_efficiency}
\end{figure}

For the purpose of understanding the results presented in this paper, it is particularly important to highlight the effect of the finite observation window. For any transit mission, the probability of detecting small planets decreases for long orbital periods due to the combination of geometric transit probability and decreased number of transits  occurring within the duration of the mission. Given the small number of such planets, as well as the increased difficulty in distinguishing planets from false positives for periods approaching one year due to Kepler's $\sim 3.5$ year primary mission, this study focuses on planets with orbital periods of less than 200 days.

To simulate the detection biases of the Kepler mission, we take into account both the geometric transit probability and the detection efficiency of the Kepler pipeline. We use the following steps to obtain a simulated catalogue of ``observed'' planets that may be directly compared to the Kepler catalogue.

\begin{description}[leftmargin=2em,style=sameline]
\item[Step 1] For each model (Table \ref{tab:threemodels}) we perform 200 N-body simulations. Each simulation begins with the same set of initial $(a,e,I)$, but all other orbital elements $(\Omega, \omega, \lambda)$ are initialized randomly. Each planet is assigned a radius following the core radius model of \citet{Zeng_2016}, assuming a rocky planet with 30\% iron.

\item[Step 2] For each run, we ``observe'' the final simulated planetary system from 2,000 different viewing angles distributed randomly across the sky. We select each viewing angle that results in least one transiting planet.

\item[Step 3] We use ExoplanetsSysSim\footnote{\url{https://github.com/ExoJulia/ExoplanetsSysSim.jl}} \citep{Hsu_2019} to compute the detection probability of each planet, given that the planet transits. To compute the detection probability of each planet, we use the corresponding transit depth and number of transits to compute the signal and draw the level of noise (due to a combination of photon noise, umodeled instrumental variability and stellar variability) based on drawing the properties of a random G star that was surveyed by Kepler (after applying the same cuts to filter for main-sequence dwarfs that are unlikely to be binary stars; see \citet{Hsu_2019}). The detection efficiency model uses the one-sigma depth function and window function interpolated to the observed orbital period and transit duration. Planet detection probabilites for two sample G dwarfs are shown in Figure \ref{fig:det_efficiency}.

\item[Step 4] We record the detection probability of each permutation of transiting planets. If an observation angle gives $N$ transiting planets, we record all $2^N - 1$ possible detection combinations and their respective probabilities.
\end{description}

\noindent To better understand how probabilities are recorded in Step 4, consider an example observation that results in $N=2$ transiting planets, called planets A and B. Let $p_{\rm A}$ and $p_{\rm B}$ be their respective detection probabilities. In this instance, we would record three possible transit events:

\vspace{1em}

\begin{tabular}{rcl}
  Event & $N_{\rm transits}$ & Event frequency \\
  \hline
  only A detected    & 1 & $p_{\rm A}(1 - p_{\rm B})$ \\
  only B detected    & 1 & $p_{\rm B}(1 - p_{\rm A})$ \\
  A\&B detected & 2 & $p_{\rm A} \, p_{\rm B}$
\end{tabular}

\vspace{1em}

\noindent The number of possible transit events for a system with $N$ transiting planets is given by $2^N - 1$, and we record the probability of each possible event per observed system with at least one transiting planet.

This process results in a simulated catalogue of observed planets that is weighted by the probability of detection averaged over all viewing angles. The resulting simulated catalogs accurately models the biases that sculpt the Kepler catalogue so that comparisons between simulations and the Kepler catalogue are meaningful.

%
%
\section{Results}
\label{sec:results}

\subsection{Solid mass as a key parameter}
\label{sec:results:mass}

\begin{figure*}
  \centering
  \includegraphics[width=0.7\textwidth]{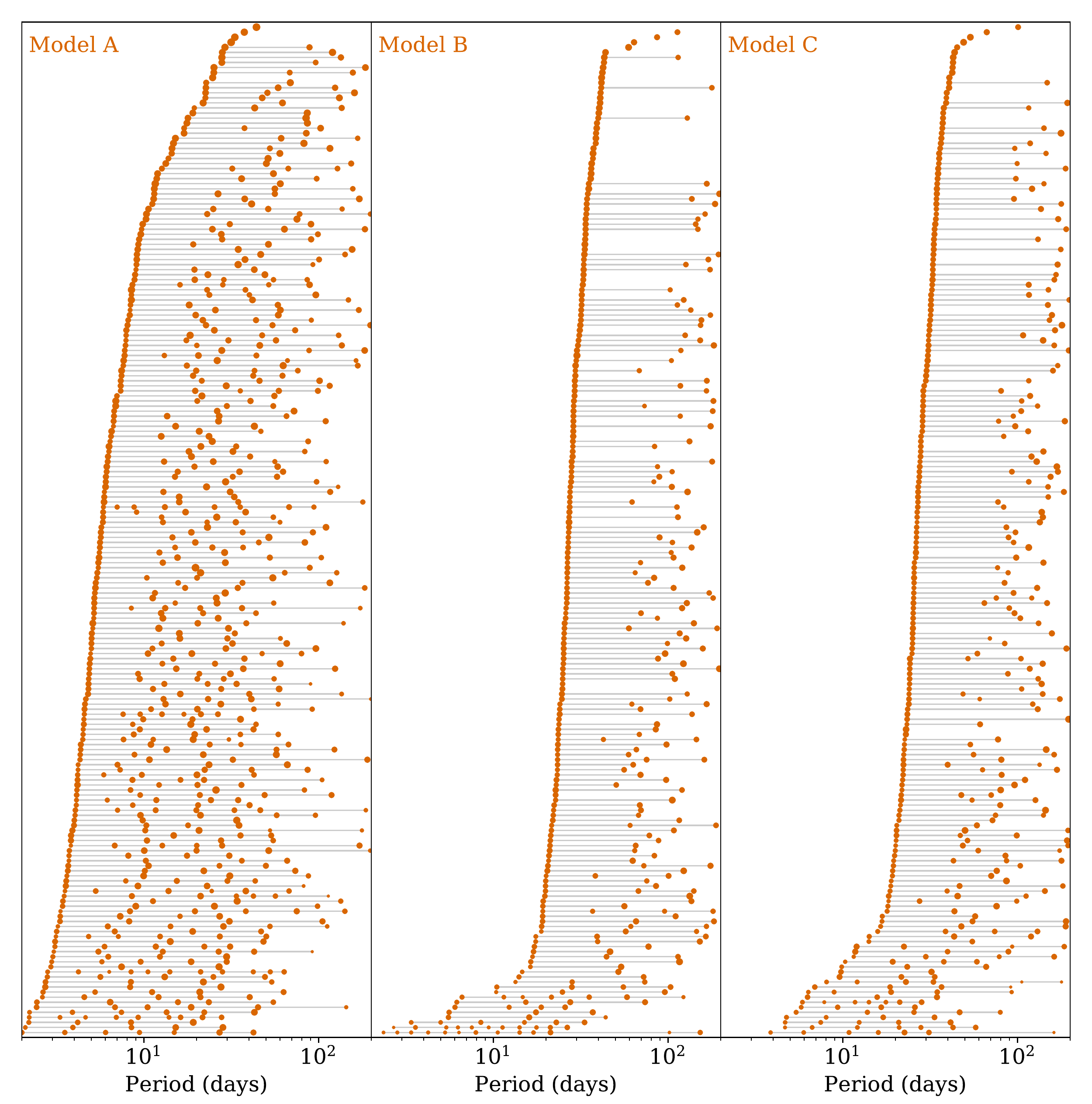}
  \caption{Overview of all the planetary systems formed in each of our models (Table \ref{tab:threemodels}). Each planet is shown as a circle proportional to the planet's core radius and each planetary system is shown in a different row. We only include planets with $P < 200$ days. For systems with at least 2 planets inside 200 days we draw a horizontal line from the inner to the outer planets. Models B and C produce comparable sets of planetary systems with similar transit multiplicities, while model A differs significantly, highlighting the impact of total solid mass on formation outcomes.}
  \label{fig:lovis_all}
\end{figure*}

Models A, B, and C were designed to be able to test the effects of three parameters: $\Sigma_0$, $\gamma$, and $M_{\rm total}$ (Table \ref{tab:threemodels}). Models B and C, which had the same total embryo mass but different $\Sigma_{0}$ and $\gamma$ values, produced nearly indistinguishable transit multiplicities, highlighting the importance of total solid mass as a key parameter for formation outcomes (which has also been highlighted in other studies, e.g., \citet{Dawson_2016} and \citet{MacDonald_2020}). For the remainder of the paper we compare only models A and B to focus on the differences between high- and low-mass systems.

\subsection{Migration traps and planet clusters}
\label{sec:results:clusters}

\begin{figure*}
  \centering
  \includegraphics[width=0.9\textwidth]{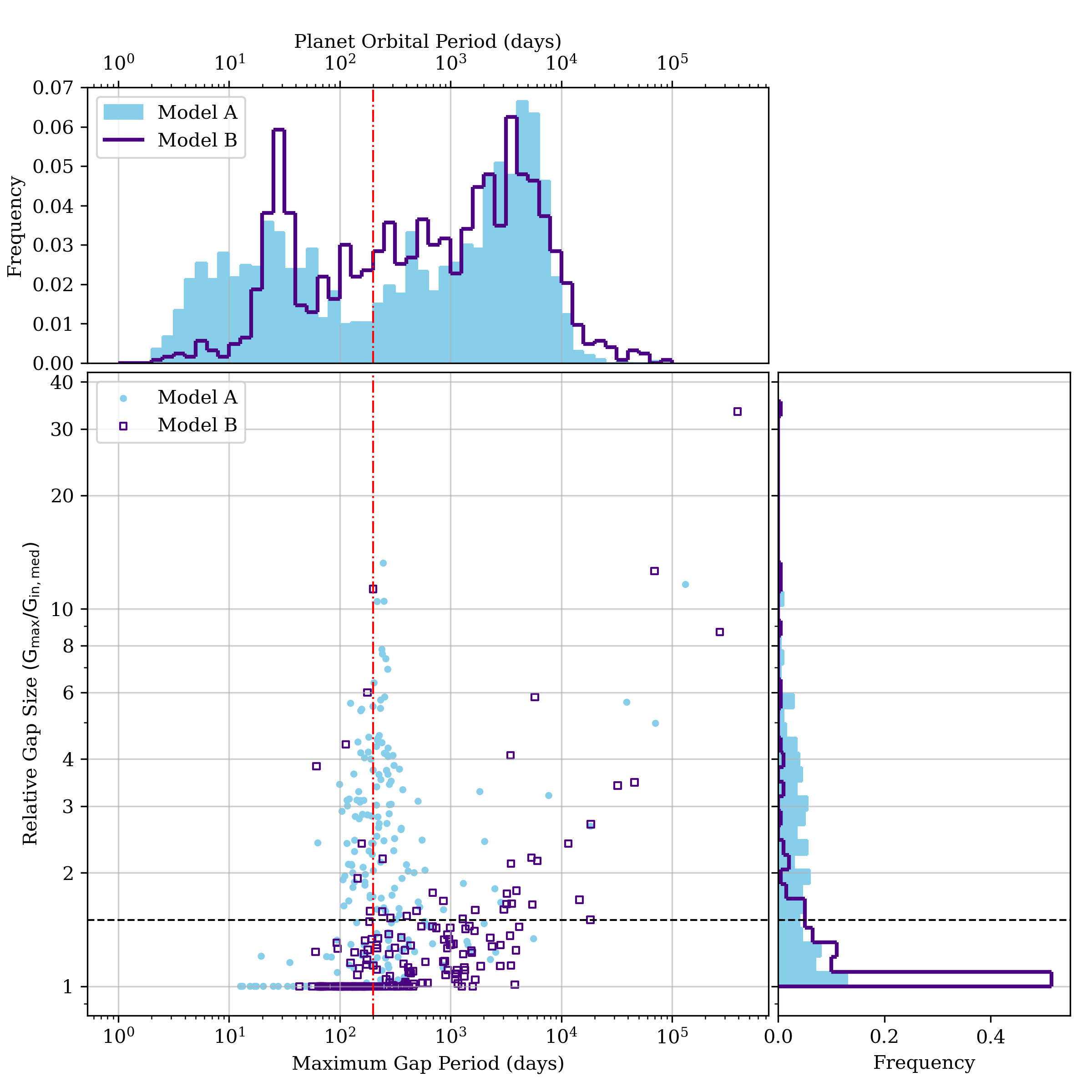}
  \caption{Identifying whether clusters are present in models A and B. The scatter plot shows the location of the largest gap in a system (here, defined as the mean period of the two planets with the largest period ratio) with respect to relative gap size. Relative gap size is defined as $G_{\rm{max}}/G_{\rm{in,med}}$, where $G_{\rm{max}}$ is the largest period ratio in a system, and $G_{\rm{in,med}}$ is the median period ratio of all planets interior to and including the planets that define $G_{\rm{max}}$. A relative gap size of 1 means that the largest gap in the system occurs between the two innermost planets in the system. We only include bodies more massive than $0.2 M_{\oplus}$ (equal to the largest initial embryo mass) so that the analysis of clustering does not include embryos that avoided collision. The top histogram shows where planets are located in each model, and the right histogram shows the distribution of relative gap sizes in each model. The red dash-dot line at 200 days roughly matches Kepler's completeness limit. The black dashed line is placed at a relative gap size of 1.5, dividing the sample into ``large gap'' and ``small gap'' systems. Model A resulted in 129/200 systems with a maximum gap ratio greater than 1.5 (large gap), while model B only resulted in 36/200.
  }
  \label{fig:clustering}
\end{figure*}

Figure \ref{fig:lovis_all} shows the final periods of all the planetary systems formed in our simulations. We have sorted the systems by the period of the innermost planet and added horizontal lines from the inner planet to the outer. Note that in both the Kepler database and our simulations we exclude all planets beyond 200 days; at longer periods the occurrence rate is poorly characterised due to the reduced geometric transit probability, smaller number of transits during the Kepler mission, and increased concerns about contamination of Kepler's planet candidate catalog \citep{Hsu_2019}. However, all of the systems plotted in Figure \ref{fig:lovis_all} have additional planets beyond 200 days. We warn the reader that Figure \ref{fig:lovis_all} makes it appear that the innermost planet pair is more widely spaced than the other planet pairs in a given system. This is an optical illusion caused by sorting the systems by the period of the inner planet. We have verified that the period ratio distribution for the innermost pair and the second innermost pair are indistinguishable.

We characterize the occurrence of clustering in each model by examining whether a large gap exists in a given planetary system. We define the relative gap size in a system, $G_{\rm{max}}/G_{\rm{in,med}}$, as the largest period ratio in a system ($G_{\rm{max}}$) divided by $G_{\rm{in,med}}$, the median period ratio of all planets interior to and including the planets that define $G_{\rm{max}}$. Systems with no clustering should have a relative gap size close to 1, while systems with a clearly defined inner and outer cluster will have a larger relative gap size. In order to compare clustering in each model, we select 1.5 as the gap ratio cutoff. Model A resulted in 64.5\% of systems (129/200) with a maximum gap ratio greater than 1.5 (large gap, indicative of clustering), while model B only resulted in 18\% of systems (36/200) having evidence of clustering. This is shown in Figure \ref{fig:clustering}, which shows the location and size of gaps for systems in models A and B, excluding embryos that did not experience any collisions. Here, the maximum gap period is defined as the mean period of the two planets with the largest period ratio. The scatter plot shows that model A tends to form systems with larger relative gap sizes than model B, with most of the largest gaps centered around 200 days.
The rightmost histogram shows the distribution of relative gap sizes in models A and B. Model A frequently forms systems with high relative gap sizes, indicative of a clear inner and outer cluster. Model B systems form distinct clusters less frequently, as shown by the strong peak in the histogram around a relative gap size of 1. The lack of clustering in model B systems is related to the observed dichotomy in transit multiplicities, as systems that do not form clusters are less likely to have multiple Kepler-detectable planets.

Model A simulations experience more disc migration due to the higher total mass, frequently pushing planets past the inner edge of the disc ($P \sim 10$ days). The lower-mass runs (model B), on the other hand, very rarely migrate that far. The top histogram in Figure \ref{fig:clustering} highlights the lack of short period model B planets compared to model A. Model A exhibits a bimodal distribution of planet locations, with the lack of planets corresponding to the large gaps near 200 days. While model B also shows a bimodal distribution of planet locations, each peak is less well-defined compared to model A, and there are not a significant number of large gaps that correspond to this region. This indicates that the migration traps seen in the high-mass simulations (model A) frequently cause systems to form distinct inner and outer planet clusters. There is evidence of this clustering in observational data; \citet{Millholland_2022} recently found that the ``edge-of-the-multis,'' the outer edge of Kepler's STIPS, tends to occur around orbital periods of 100-300 days.

\begin{figure*}
  \centering
  \includegraphics[width=0.97\textwidth]{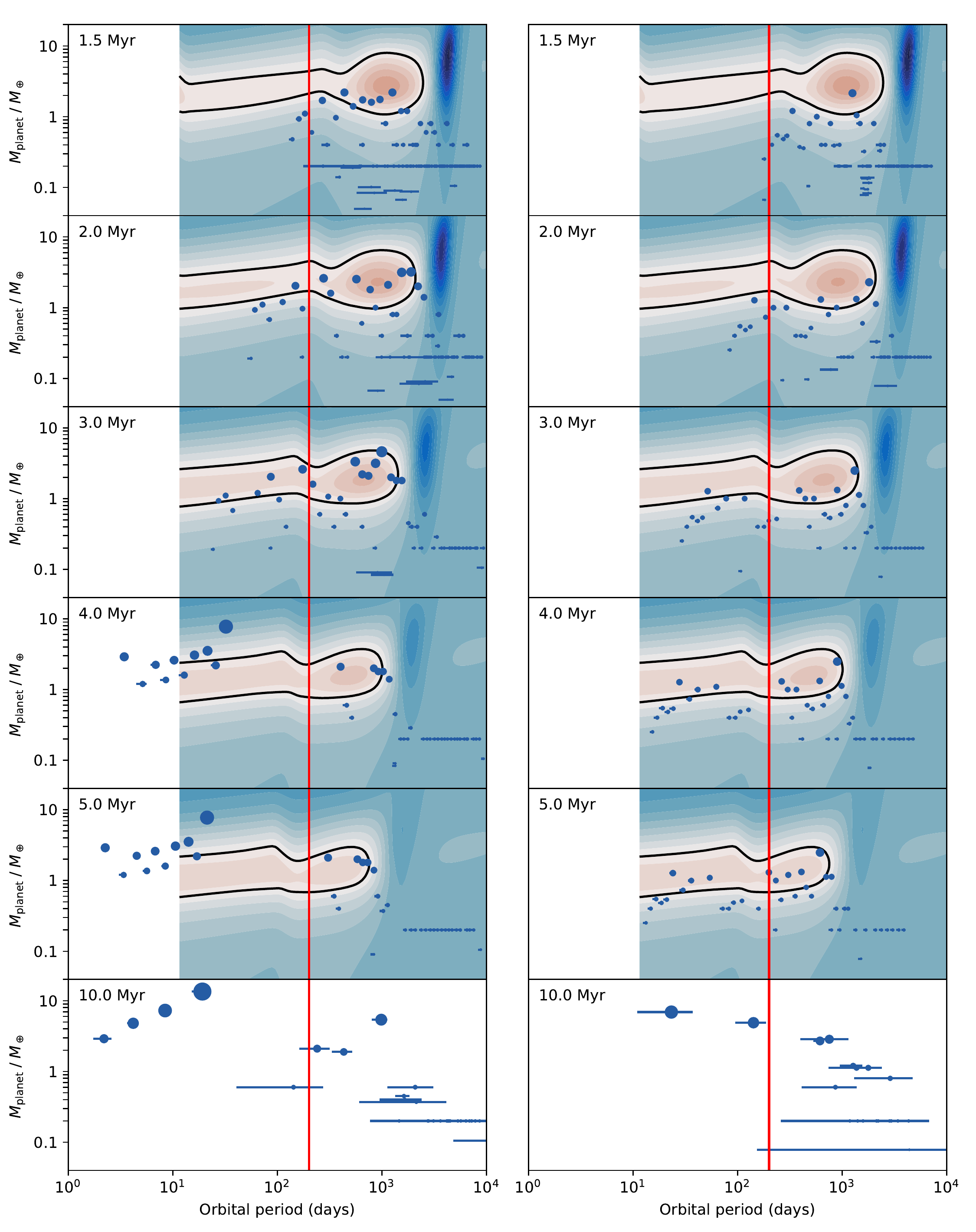}
  \caption{Snapshots from a sample run from both our high-mass model A (left) and low-mass model B (right). Planetary embryos are shown as circles with size proportional to their rocky core radius. The vertical line, at 200 days, roughly matches Kepler's completeness limit. Between 3 and 4 Myr, the high-mass model (left) sees rapid growth that takes the inner embryos above the migration trap (black curve). Freed from the trap, the inner embryos migrate quickly and become dynamically separated from the outer embryos. This helps the inner planets remain dynamically colder after the disc phase. In the low-mass model (right) all embryos remain together and form a single dynamically hot cluster. See Figure \ref{fig:lovis15} for typical outcomes.}
  \label{fig:snapshots}
\end{figure*}

\begin{figure*}
  \centering
  \includegraphics[width=0.98\textwidth]{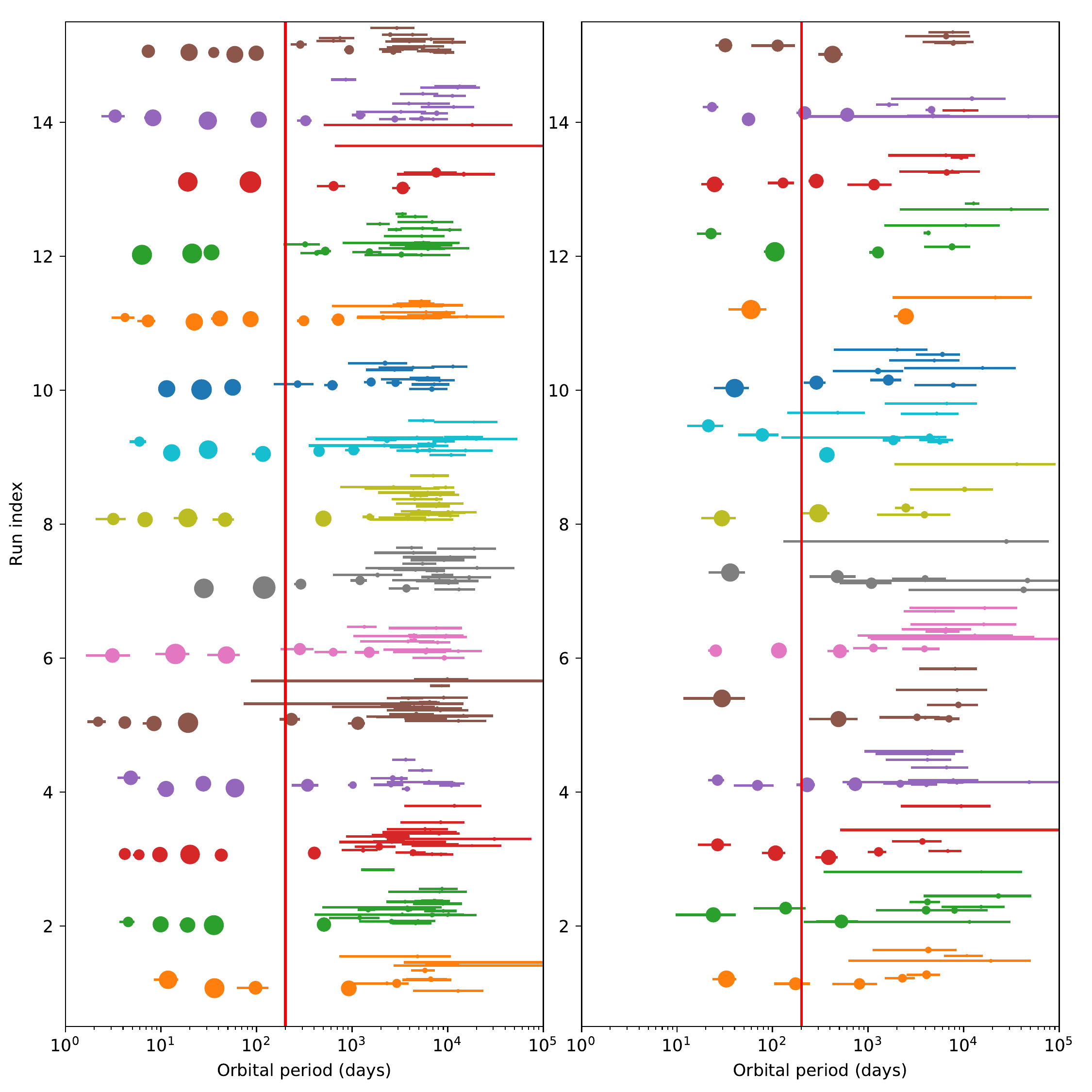}
  \caption{Final configurations for 15 sample simulations from our high-mass model (model A, left) and our low-mass model (model B, right). Each system is shown in a different colour. Planets are shown as circles proportional to their rocky radius and they are offset vertically in proportion to their orbital inclination relative to the invariant plane. The horizontal lines range from the planet's periastron to apastron. Overlapping lines point to orbit crossings and likely future collisions among the outer planets. The vertical line roughly matches Kepler's completeness limit. The high-mass model (left) often produces migration traps which lead to two distinct planet clusters. The inner cluster typically remains relatively cold, compact, and isolated from a much hotter outer cluster. Figure \ref{fig:snapshots} shows the formation history that leads to separated clusters in one model but not the other.}
  \label{fig:lovis15}
\end{figure*}

We find that the low-mass runs (model B) are less stable than our high-mass runs (model A), experiencing more collisions and producing systems with higher eccentricities and inclinations. At first this may appear to differ from a common rule of thumb that more massive systems are more likely to be dynamically excited \citep[e.g.][]{Chambers_1996,Faber_2007}. However, this conventional wisdom is based on simulations that initialized planets on random initial orbits and do not consider the effects of migration and planet traps.

Figure \ref{fig:snapshots} explains this result: the low-mass runs (model B; right column) typically form smaller planets that become caught in the migration trap produced by co-rotation torques. When this happens, the simulation typically produces a single cluster of planets that quickly becomes dynamically unstable. Our high-mass runs (model A; left column) frequently form several inner planets above the mass scale where co-rotation torques are effective, causing the inner planets to quickly migrate inward and split off from the outer planet, ultimately resulting in two distinct planet clusters that are largely dynamically isolated. When this happens, the outer planet cluster still becomes unstable, but the inner cluster often appears mostly unaffected by the dynamical havoc in the outer system.

The runs in Figure \ref{fig:snapshots} are chosen to highlight the difference between the formation of one planet cluster vs. two. However, the typical simulation results are as not clear cut. Figure \ref{fig:lovis15} shows a representative sample of the planetary architectures typically produced by our two models. The general trend is that the high-mass model (model A; left column) (a) often, \textit{but not always}, produces two distinct planet clusters with a visible gap between them; (b) produces $\sim$4 planets inside 200 days (versus $\sim$2 for the low-mass model); and (c) is more likely to produce compact close-in dynamically cold systems. We speculate that close-in compact Kepler systems may be comprised of planetary systems that separate into two distinct dynamically isolated clusters.

\subsection{No dichotomy in system architectures}
\label{sec:results:inclination}

\begin{figure*}
  \centering
  \includegraphics[width=0.49\textwidth]{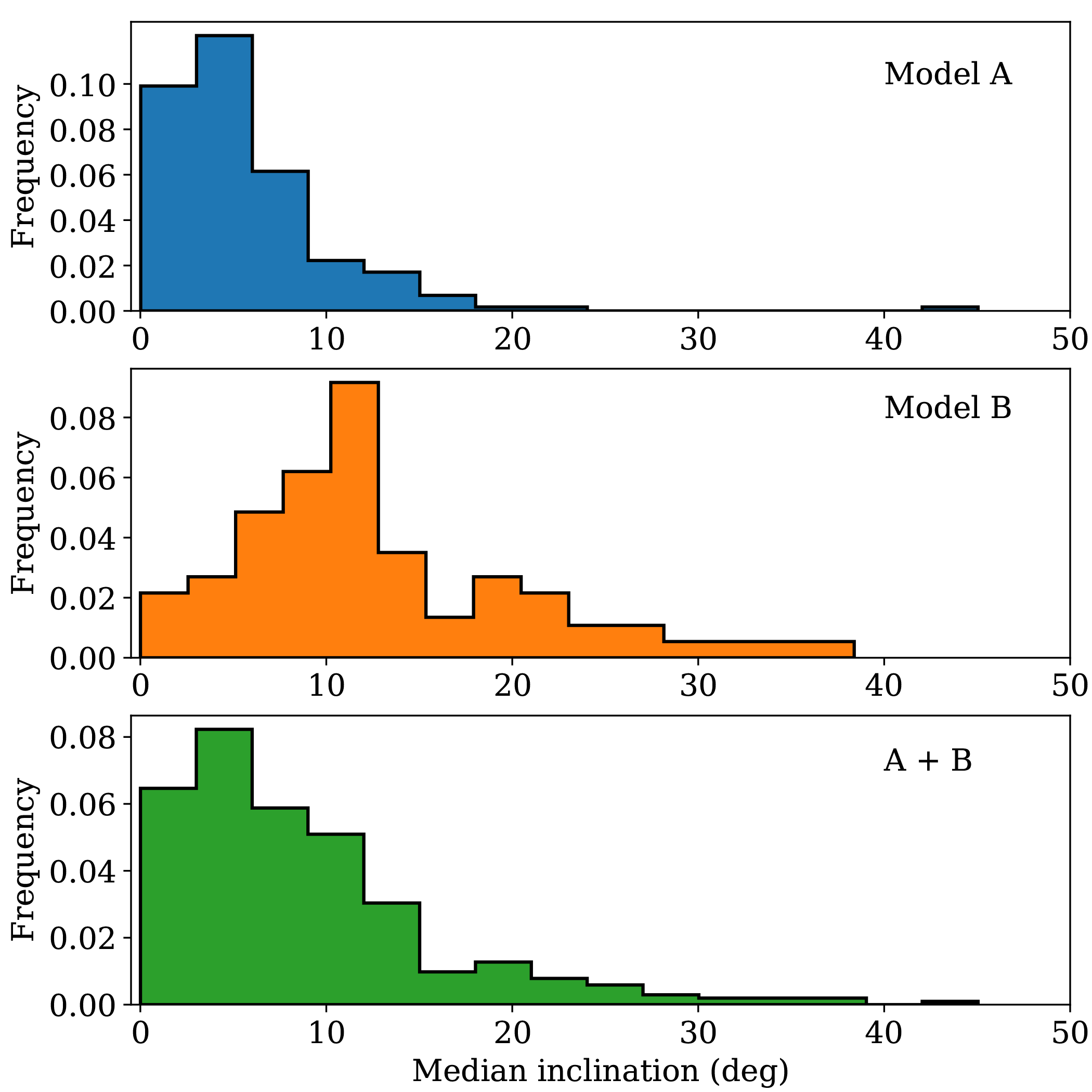}
  \includegraphics[width=0.49\textwidth]{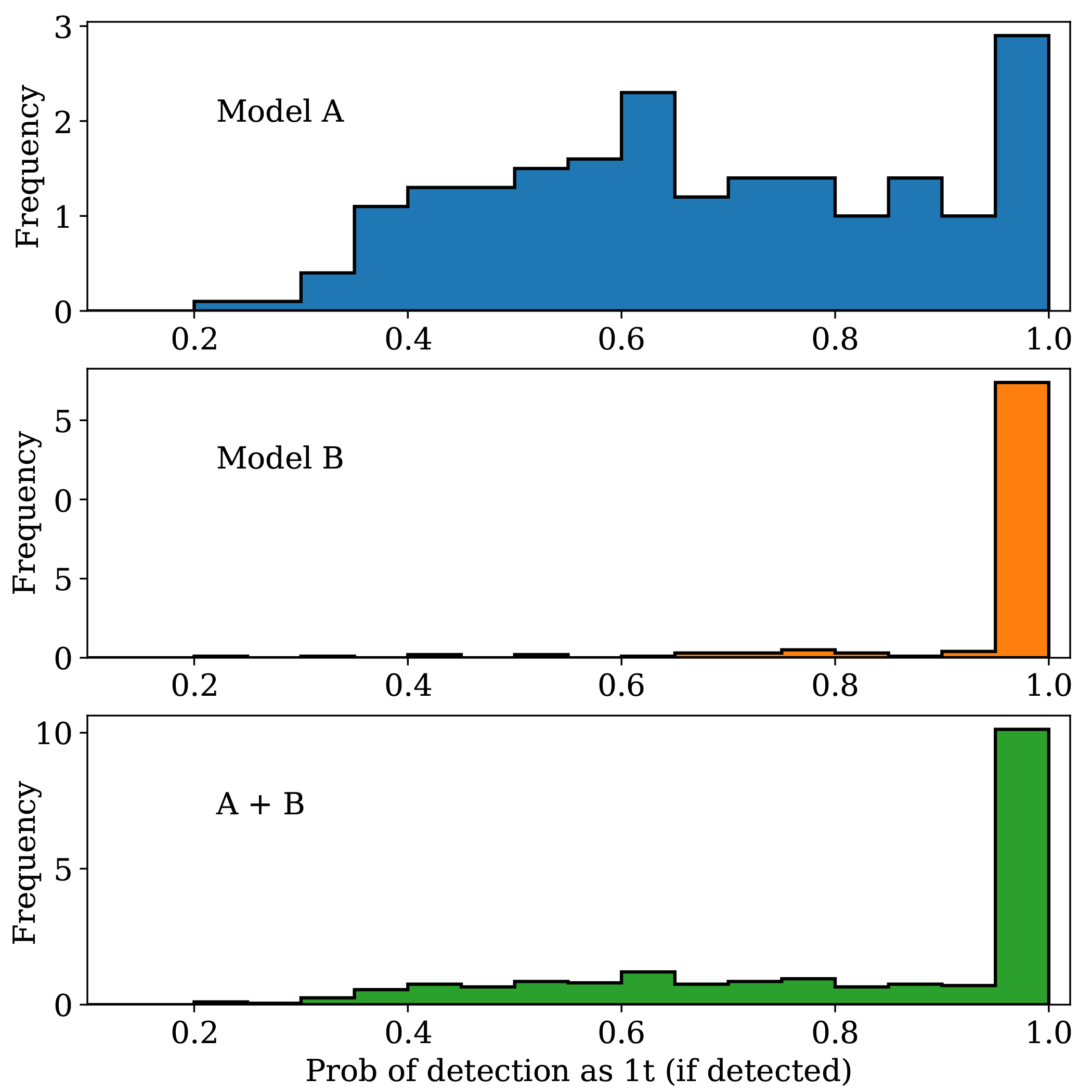}
  \caption{\textit{Left:} Distribution of median inclinations between adjacent planet pairs for each planetary system in models A and B. Only planets within 200 days are included. Simulations that formed only one planet inside 200 days are excluded from those plots. \textit{Right:} Histogram of the probability that a planetary system, if detected, will be identified as a single-transit system. The bottom histograms, labelled ``A+B'', are built by adding the histograms from models A and B. In all cases, the main take-away is that neither model A or B or their sum shows any clear bimodality in the mutual inclinations, suggesting that a bimodal distribution of mutual inclinations is unlikely to be the driving cause of the dichotomy in single-transit frequency.}
  \label{fig:dichotomy-hist}
\end{figure*}

Many authors have argued that the Kepler dichotomy points to inherent differences in system architectures in which the Kepler sample is composed two exoplanet populations, one population that is dynamically hot and characterised by high mutual inclinations, and one population that is dynamically cold \citep{Johansen_2012,Moriarty_2016,Izidoro_2017}. Some authors have estimated the sizes of these populations \citep{Moriarty_2016,Izidoro_2017,He_2019}, and some have identified these supposed populations with dynamical instabilities \citep{Johansen_2012,Izidoro_2017}. 

When we started this investigation, one of our primary goals was to better understand the nature of these ``hot'' and ``cold'' populations, taking into account the effect of Kepler detection biases as well as any correlations that arose from the planet formation process. However, we find that planet formation models \textit{do not} seem to produce drastically different formation histories. Dynamical instabilities are not a simple on/off process; there is a smooth continuum of simulation outcomes from systems that experience virtually no dynamical excitation after the disc dissipates to those that have strong interactions, and everything in between. 

The mutual orbital inclinations of the final planetary system has a substantial effect on the probability that a system is observed to have multiple transiting planets. While the final planetary systems have a range of mutual inclinations, we find no apparent bimodality in their mutual inclinations or formation histories.
Figure \ref{fig:dichotomy-hist} shows some of our attempts to search for any potential bimodality in the mutual inclination distribution. The left plots show the distribution of median mutual inclination for each system in models A and B. In both cases, the histogram shows a wide distribution of outcomes with only one peak. While models A and B clearly have their peaks at different places, the distribution of median inclinations is sufficiently broad that the combined model does not end up having two peaks, but instead the combination effectively has a single peak somewhere in the middle. Though not shown, we also looked for bimodalities in the minimum and maximum inclinations and other architectural features.

Finally, we considered the possibility for a dichotomy in the classification of systems as single- or multiple-transiting systems, even in the absence of a bimodality in the inclination distribution. Indeed, we did find evidence of a dichotomy in the probability of Kepler detecting a single transiting planet, with model B systems being much more likely to be identified as a single transiting system than model A systems. This is related to the dichotomous relative gap distribution in models A and B, where model B gap sizes are indicative of infrequent formation of inner clusters with multiple Kepler-detectable planets compared to model A (see the rightmost histogram in Figure \ref{fig:clustering}). Note that the distribution of relative gap sizes is not equivalent to ``the Kepler dichotomy'' (as used in the literature), but the relative gap sizes of systems generated from a mixture of models A and B could be considered dichotomous because the resulting distribution is best described by two distinct populations, i.e. systems that form a compact inner cluster reminiscent of Kepler STIPs and systems that do not. The right panels of Figure \ref{fig:dichotomy-hist} show histograms of the probability that a planetary system will be classified as a single transit. There is a large excess of systems with Prob(1t) $\approx$ 1. In the ``A+B'' plot, 60\% of the peak is caused by two-planet systems with a mutual inclination above 5.8$^\circ$ (median around 12.6$^\circ$), a further 33.7\% are systems with only one planet inside 200 days, and the remaining 6.3\% comes from 11 three-planet systems with more complex orbital configurations. Note that the mutual inclinations of these two-planet systems are not exceptionally high compared to the overall distribution. However, in combination with a low inherent planet multiplicity, they seem to drive a peak in single transit detections.

\subsection{The cause of the Kepler dichotomy}
\label{sec:results:dichotomy}

We compare our the transit multiplicities of models A and B to a sample of Kepler planets. The sample contains Kepler planet candidates with disposition scores $>0.5$, radii of $1-3 R_{\oplus}$, and periods $<200$ days around G dwarf stars (i.e. $5300 \;\rm{K} < T_{\rm{eff}} < 6000 \;\rm{K}$). We find that the low-mass model (model B) produces a greater excess of systems with only one transiting planet (i.e. an exaggerated Kepler dichotomy) than model A or observed in the Kepler sample. A roughly equal mix of models A and B closely mimics the relative frequencies of systems with one, two and 3+ planets detected by Kepler.
More specific values for observed occurrences of single, double, or 3+ transiting systems are shown in Table \ref{tab:nodal_alignment}.

\begin{figure*}
  \centering
  \includegraphics[width=0.97\textwidth]{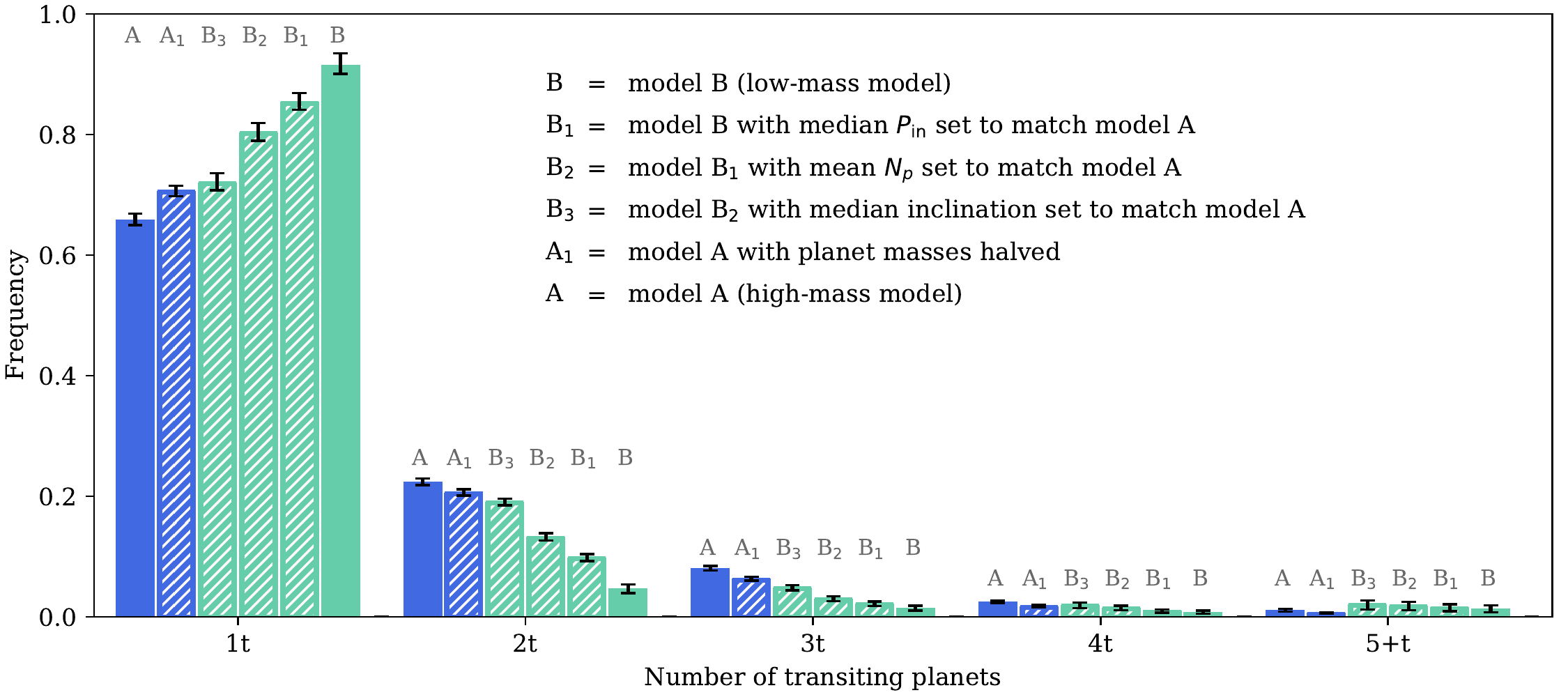}
  \caption{Frequency of single (1t) and multi-transit (2t, 3t, etc.) systems for models A and B, as well as four thought experiments. The low-mass runs (model B) produce planets with larger periods, wider separations, higher mutual inclinations, and smaller planet radii. The thought experiments progressively alter each of these variables. In $B_1$ we scale down planet periods so that the innermost period matches model A. In $B_2$ we also scale down the separations to make the number of planets inside 200 days match model A. In $B_3$ we also lower the inclinations to match model A, and in $A_1$ we lower planet masses in half, similar to model B. Since $A_1$ and $B_3$ look similar, these four variables explain most of the difference between the high-mass and low-mass systems. The error bars show the 25th and 75th percentiles of the distribution resulting from 10,000 bootstrap resamples.}
  \label{fig:thought}
\end{figure*}

In this section we take a deeper look at which planetary system properties are most responsible for the fraction of single-transit systems. To do this, we conduct a series of thought experiments, which we summarize in Figure \ref{fig:thought}. The figure shows the distribution of transit multiplicities in models A and B (columns labeled ``A'' and ``B'' respectively), as well as for the four thought experiments:

\begin{description}[leftmargin=2em,style=sameline]
\item [Experiment ${\bf A}_1$ -- Effects of Detection Probability] 
We took the simulation results of model A, multiplied all the planet masses by half, re-computed the planet radii, and then simulated Kepler observations. Smaller planets have a lower transit depth, and thus a lower detection probability. But there is a subtle effect here: When the probability of planet detection decreases, a higher percentage of detected systems will appear to be single-planet systems. To understand this, consider a toy model where all planetary systems have two planets with identical detection probability $p$. The probability of detecting two planets is $p^2$ and the probability of detecting exactly one planet is $2p(1-p)$. Therefore, the apparent ratio of 1-transit to 2-transit systems is $2(1-p)/p$, and as $p$ decreases, that ratio increases. Experiment $A_1$ is a rough way to estimate the size of this effect.

\item [Experiment ${\bf B}_1$ -- Effects of Migration]
We took the simulated systems from model B and multiplied their orbital periods by 0.21 in order to bring the median period of the innermost planet ($P_{\rm in}$) in line with that of model A, before simulating Kepler observations. We can think of this experiment as a rough indicator of the effect of overall migration. Greater planet migration brings more planets into the Kepler detection window (red line of Figure \ref{fig:snapshots}). In addition, close-in planets have a higher transit probability and more transits, further increasing their probability of detection.

\item [Experiment ${\bf B}_2$ -- Effects of True Number of Planets]
We took the systems from $B_1$ and multiplied all planet-planet separations (i.e. $\Delta a = a_{i+1}-a_{i}$) by 0.55 (while keeping $a_1$ constant) to make the total number of planets inside 200 days match the value for model A. Since Kepler's completeness is small for orbital periods greater than 200 days, this experiment approximates how much the intrinsic number of planets in a system impacts the observed transit multiplicity.

\item [Experiment ${\bf B}_3$ -- Effects of Mutual Inclinations]
We took the systems from $B_2$ and multiplied all inclinations by 0.53 to make the mean mutual inclination match that of model A. Previous work on the Kepler dichotomy has often focused on inclinations \citep[e.g.][]{Moriarty_2016} because planetary systems with high mutual inclinations are more likely to be observed as having only one transiting planet. This experiment estimates the effect of mutual inclinations on observed transit multiplicity.

\end{description}

We emphasize that these thought experiments are only meant to quantify the importance of these effects, which have previously been largely neglected in studies of the Kepler dichotomy, rather than fitting a specific planet formation model. Figure \ref{fig:thought} shows that these four effects (period of innermost planet, number of planets in the system with period $< 200$ days, mutual inclinations, and planet sizes) explain nearly all of the difference in the distribution of transit multiplicities between the two models presented here. For the sake of brevity, we will refer to systems with $N$ transiting planets as ``$N$-transit.'' Figure \ref{fig:thought} allows us to quantify how much each variable contributes to the difference in 1-transit frequency between models A and B.
To do this, we take the 1-transit frequency of each thought experiment model (i.e. the height of the corresponding 1t bar in Figure \ref{fig:thought}), subtract it from the 1-transit frequency of either model A or B, and divide by the difference in 1-transit frequency of models A and B. This yields a percent difference in 1-transit frequency corresponding to the factors isolated by each of the thought experiments. Because models B2 and B3 are built on models B1 and B2, respectively, we also subtract the contributions from the base models when calculating these percentages. This results in the following contributions to the difference in 1-transit frequency:

\begin{itemize}
\item $18.4 \pm 1.7$ \%  is probably due to planet sizes
\item $32.2 \pm 2.5$ \% is probably due to inclinations
\item $43.9 \pm 4.2$ \% is due to the number of planets in the system
  \begin{itemize}
    \item $19.7 \pm 1.8$ \% is due to planet separations
    \item $24.2 \pm 2.4$ \% is due to overall periods
  \end{itemize}
\item $5.5 \pm 5.8$ \% is unaccounted for (but see \S\ref{sec:results:nodes})
\end{itemize}

\noindent Uncertainties on these values and the values shown in Figure \ref{fig:thought} were computed by bootstrap resampling our catalog of simulated planetary systems, drawing 200 systems with replacement. We took 10,000 resampled catalogs and recomputed transit frequencies for each. The error bars indicate values falling into the central 50\% of the 10,000 resampled catalogs.

We emphasize that these results suggest that mutual orbital inclinations are not the primary driver of the Kepler dichotomy, at least not for our simulated planetary systems. The excess of 1-transit systems is the result of multiple correlated properties of planetary systems, where inclinations are but one component. The most important variable that determines the number of 1-transit systems in our simulations is simply the number of planets inside Kepler's detection window; this effect is responsible for nearly half of the difference between our two models. Mutual inclinations play a significant, but secondary role. Finally, there is a modest but clearly detectable effect of small-planet systems being more likely to be interpreted as 1-transit systems.

\begin{figure}
    \centering
    \includegraphics[width=\linewidth]{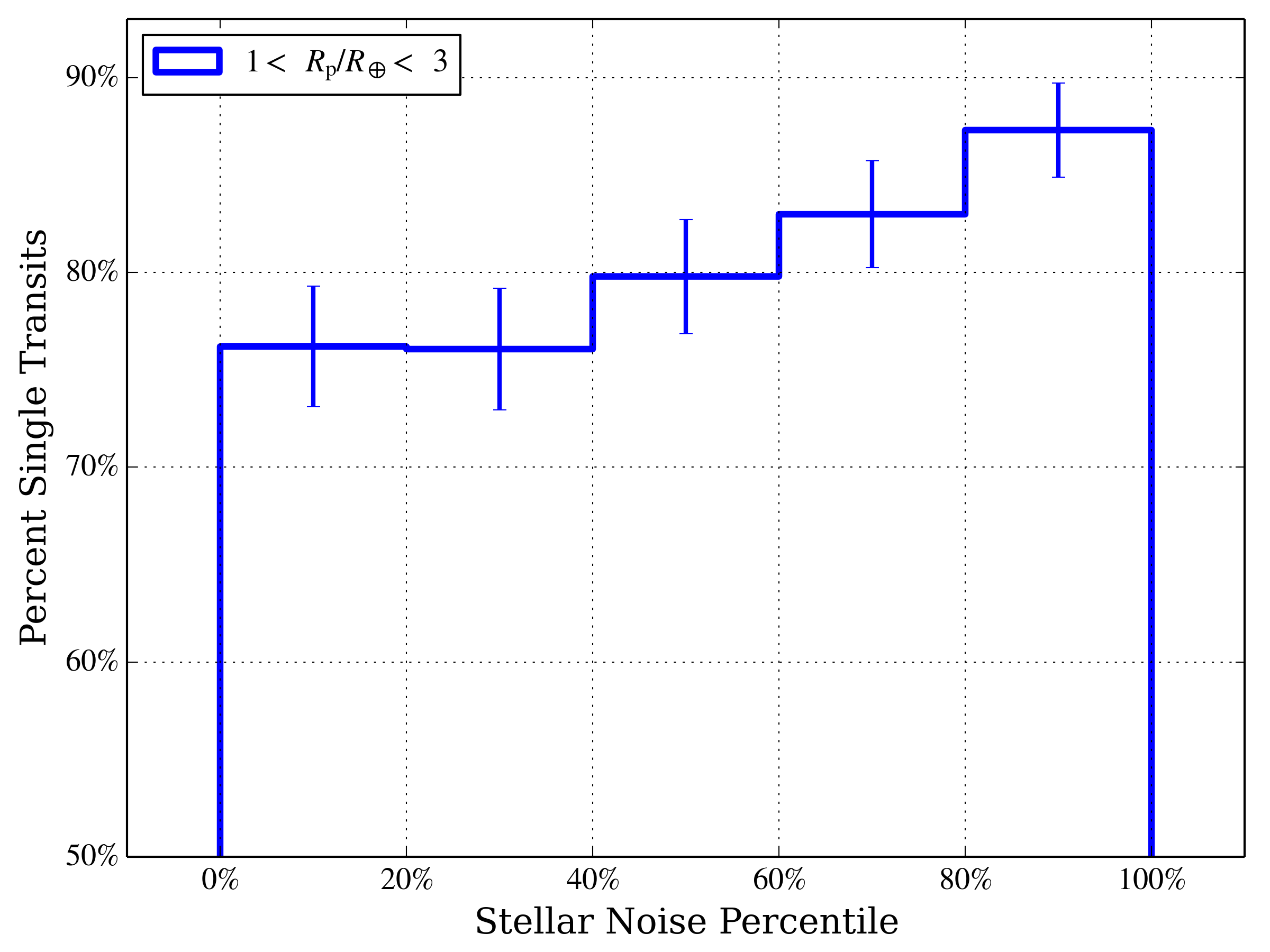}
    \caption{The occurrence of 1-transiting systems as a function of stellar noise (obtained by the 4.5 hour CDPP) for a sample of Kepler planet candidates with radii of $1-3 R_{\oplus}$ and periods $<200$ days around G dwarf stars. Error bars are shown assuming a binomial distribution. The noisiest stars tend to have larger fractions of 1-transiting systems, effectively increasing the magnitude of the Kepler dichotomy. We have verified that this trend is robust to the choice of planet radii included, especially for the noisiest 20\% of stars.}
    \label{fig:stellarnoise}
\end{figure}

However, the small-planet selection effect is not strictly about planet size --- anything that makes planets difficult to detect, such as stellar noise, will increase the magnitude of the Kepler dichotomy. We have verified that the fraction of planetary systems detected by Kepler with only one planet detected increases for Kepler targets with higher level of stellar noise, as measured by the 4.5 hour CDPP (Figure \ref{fig:stellarnoise}). This pattern holds regardless of whether we use our Kepler sample (i.e. $1-3 R_{\oplus}$ and $T<200$ days around stars with $5300 \;\rm{K} < T_{\rm{eff}} < 6000 \;\rm{K}$) or a more carefully selected sample of FGK stars that were selected based on Gaia data likely to be main-sequence G stars without significant binary contamination (see Sec 3.1 of \citet{Hsu_2019}). The pattern is also robust to the the planet size range included in the sample (we tested size cutoffs at 2, 2.5, 3, 4, and 5 $R_{\oplus}$). This demonstrates that in order to discern whether patterns are intrinsic properties of the distribution of planetary systems or purely the result of selection effects, it is important to couple formation models with a detailed model for the planet detection and vetting efficiency.

As an aside, the 6.37\% of the difference that we could not account for corresponds to 1.6\% of the ``observed'' systems. We do not think that that difference is just statistical noise since our 200,000 observations per model should be able to detect selection effects much smaller than 1\%. It is more likely that the handful of variables we have explored in this section do not cover all of the differences in the formation of a low-mass and high-mass planetary system. Incidentally, this 1.6\% effect is comparable to the effect of nodal alignment that we discuss in the next section.

\subsection{Even unstable systems show nodal alignment}
\label{sec:results:nodes}

For a planetary system with high or even moderate mutual inclinations, the only way to detect two planets is to observe the system along an angle near the line where two orbital planes meet. The only way to detect three or more planets is if the ascending nodes of the planetary orbits are aligned such that multiple orbital planes intersect near the same region in space. Therefore, the observed sample multiple planet systems are expected to show significant nodal alignment, even if the intrinsic distribution of ascending nodes is independent. In other words, the very act of selecting systems with multiple transit preferentially selects systems with aligned nodes. Most previous studies of multiple planet statistics have assumed that the ascending nodes of planets within a system are independent of each other. Over the course of this investigation, we considered the possibility that the intrinsic population of planetary systems might include planets with correlated ascending nodes and that this could play a role in the apparent Kepler dichotomy. 
We found five interesting results:

\begin{enumerate}
\item All planetary systems typically form with some amount of clustering of the ascending nodes. Even major dynamical instabilities with many collisions and drastic increases in eccentricities and inclinations seem to retain some degree of nodal alignment --- that is to say, the planet formation process results in non-uniformly distributed ascending nodes.

\item Planetary systems, in both models, where the probability of detecting three or more planets is greater than the probability of detecting exactly two, do show significantly more clustering of ascending nodes than the other planetary systems.

\item The nodal alignments seen arising from our planet formation simulations has only a modest effect on the number of systems observed to have a single transiting planet.

\item The nodal alignments that arise from our planet formation simulations appear to have a potentially significant effect on the relative frequency of many-planet systems observed in transit ($\sim14-25\%$ for systems with three or more transiting planets). If this apparent trend extends to higher multiplicities, the effects of nodal alignment may be substantial for high-multiplicity systems.

\item The amount of nodal alignment does not seem to be a distinguishing feature between our high-mass and low-mass models. It does not contribute to explaining why the low-mass model produced more 1-transit systems than the high-mass model.
\end{enumerate}

\begin{figure*}
  \centering
  \includegraphics[width=0.35\textwidth]{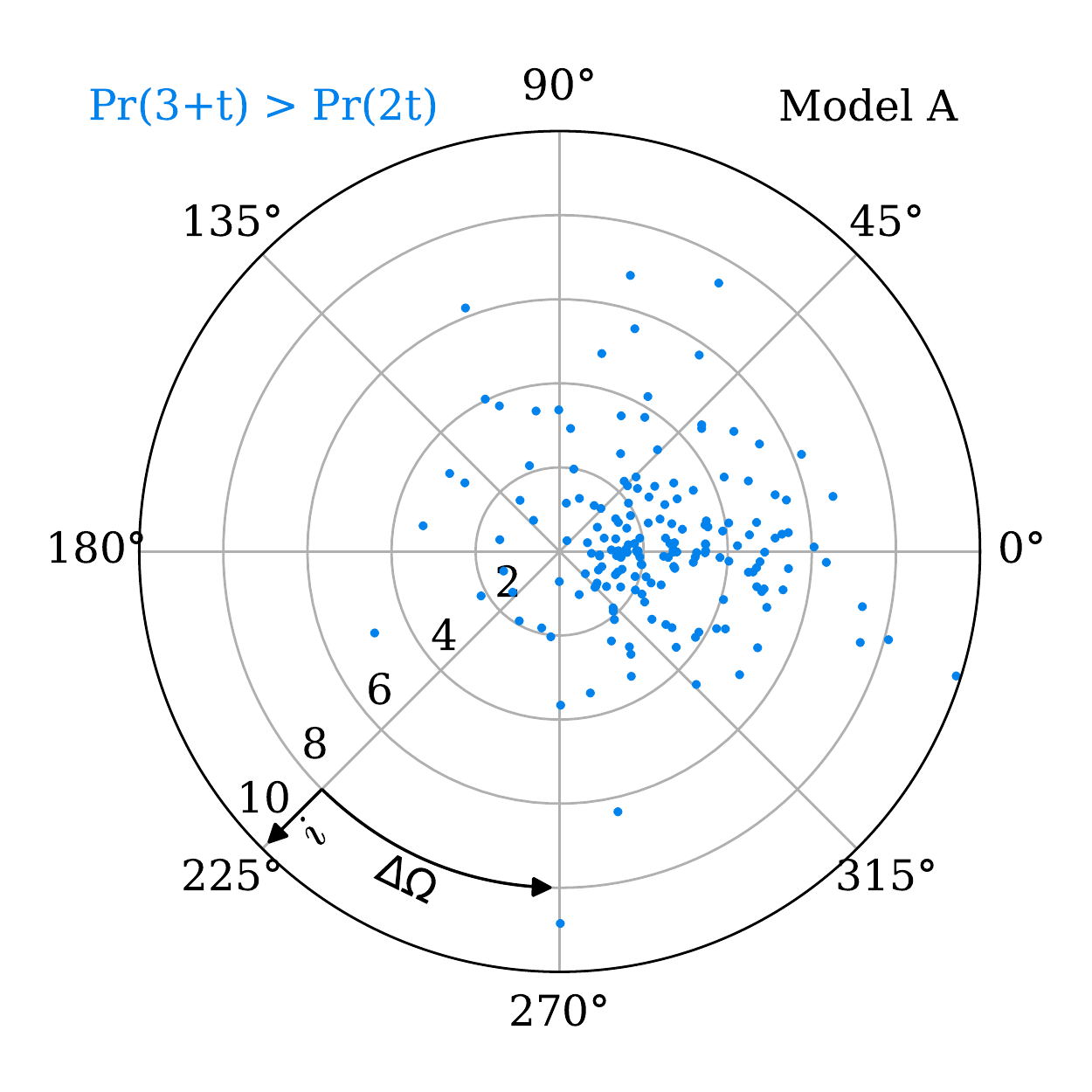}
  \includegraphics[width=0.35\textwidth]{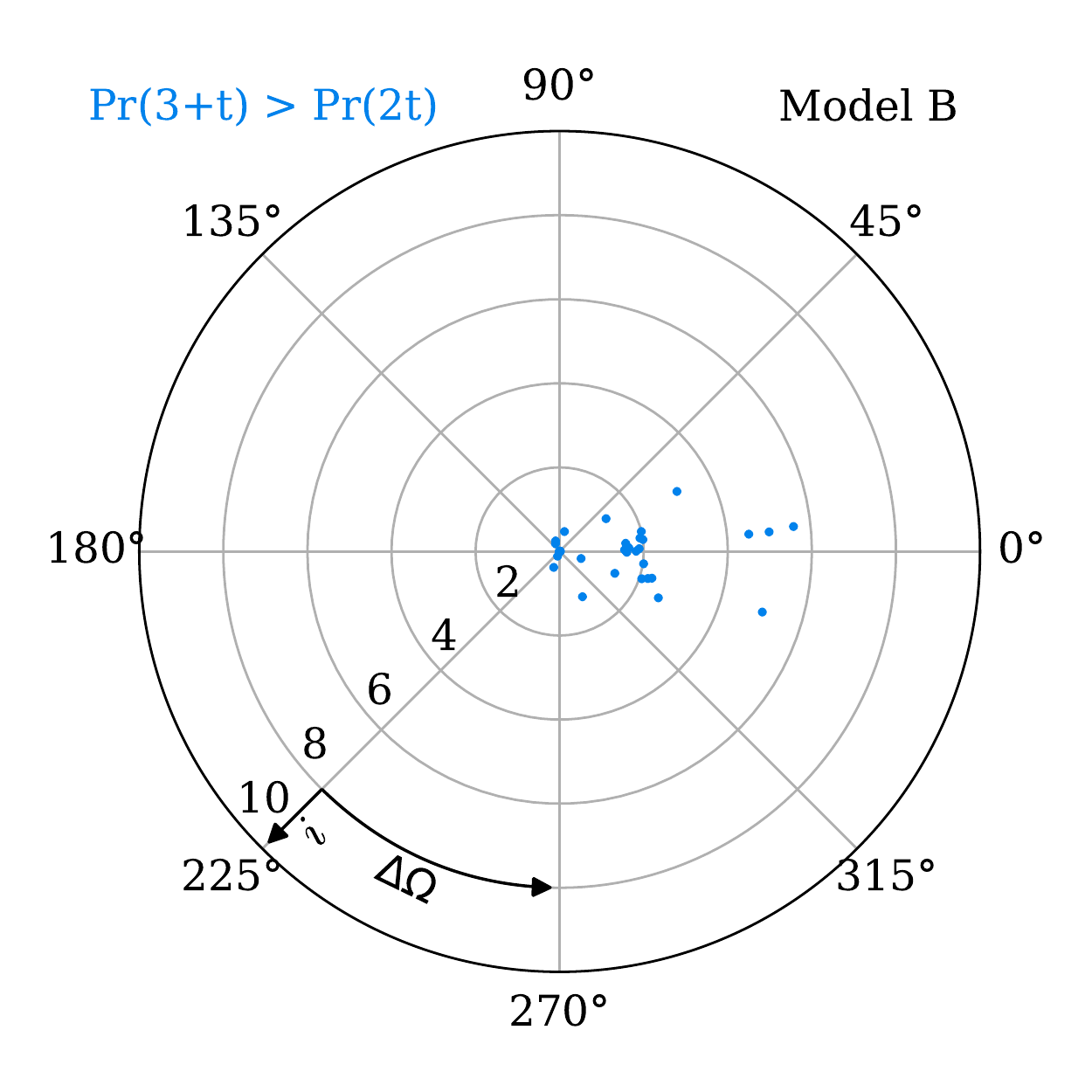}\\
  \includegraphics[width=0.35\textwidth]{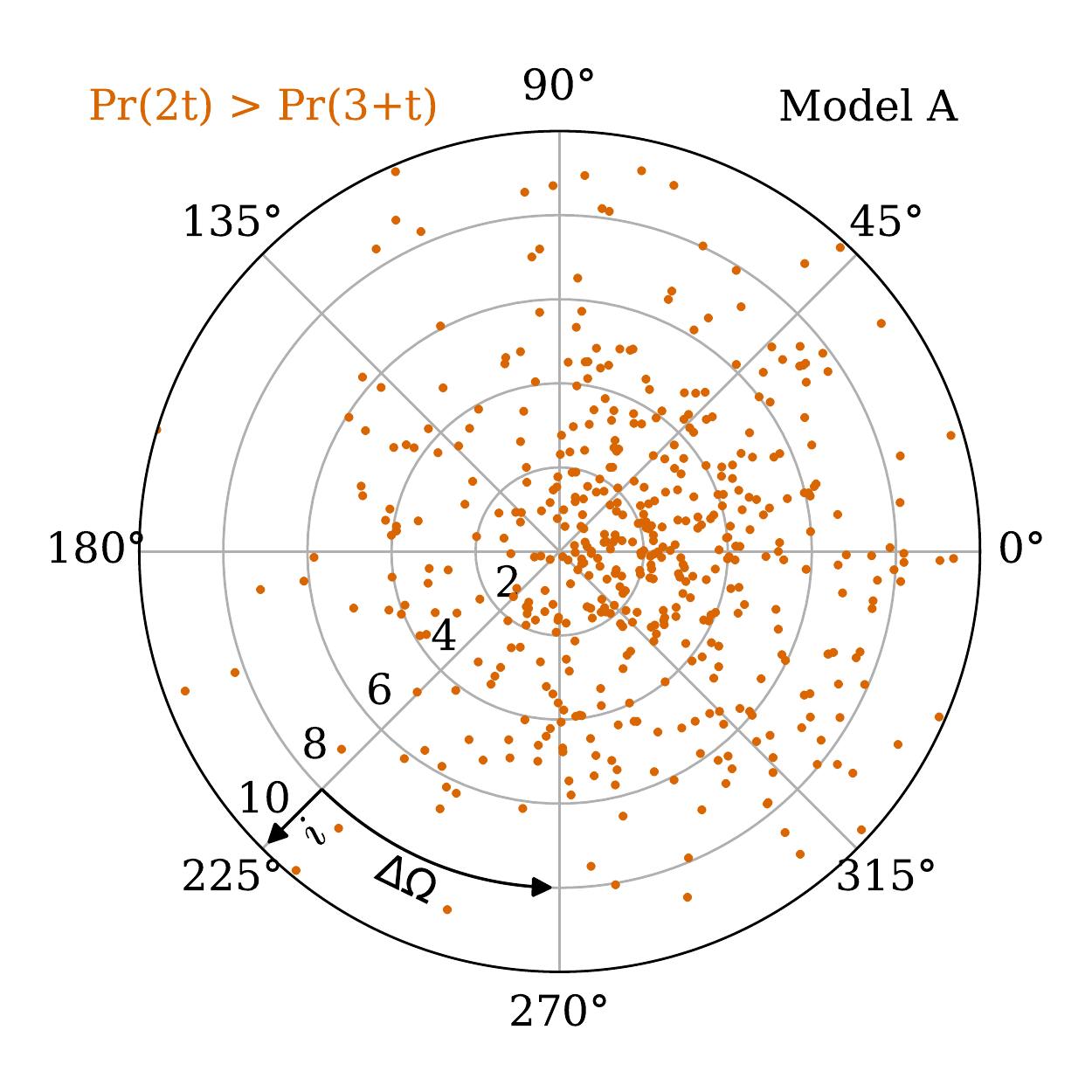}
  \includegraphics[width=0.35\textwidth]{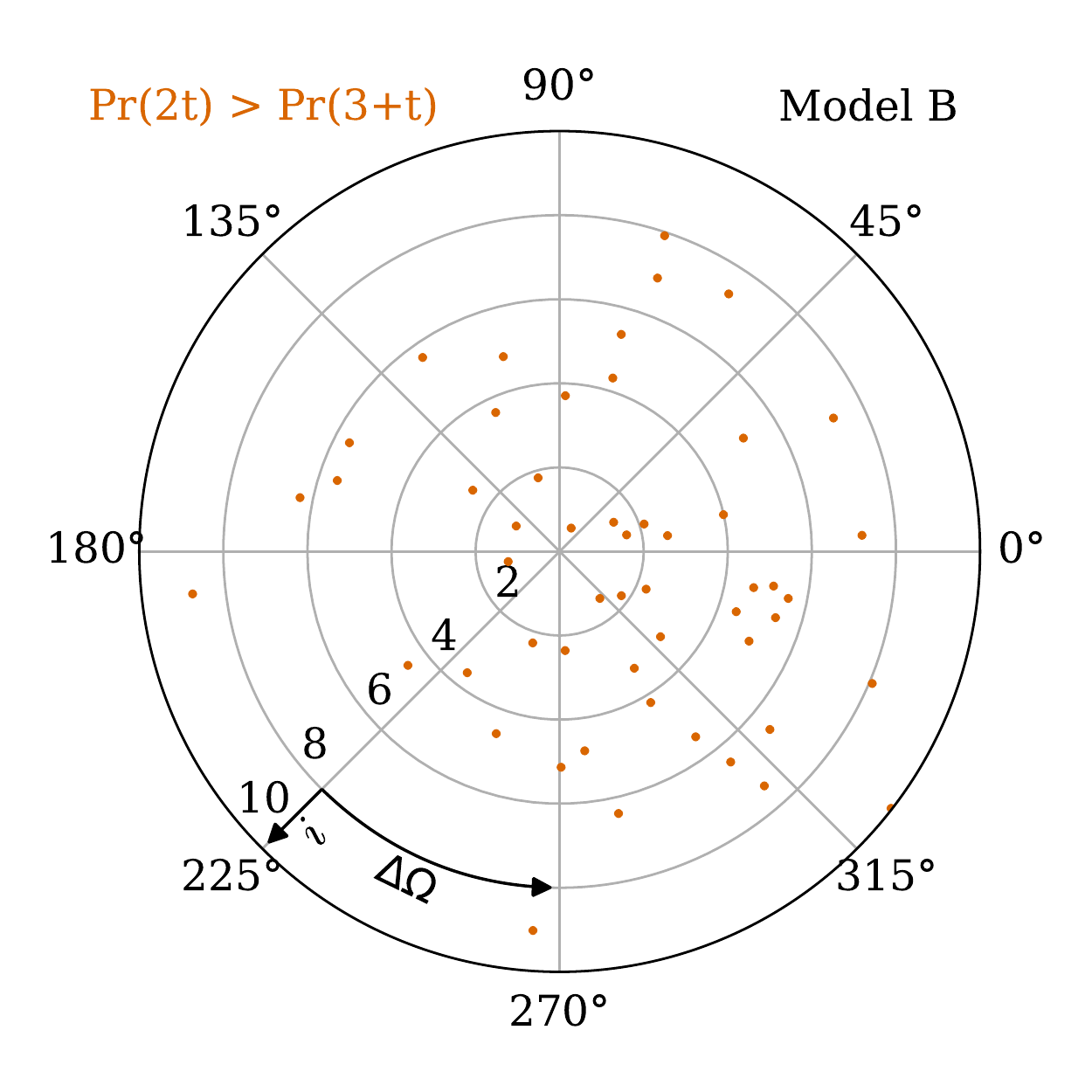}
  \caption{For each planet in a system with at least three planets inside 200 days we plot the planet's inclination $i$ relative to the invariant plane, and ascending node $\Delta \Omega$ after rotating the system so that the \textit{mean} ascending node is zero. Rotating the system like this makes nodal alignment easier to identify, as aligned nodes appear as an over-density of dots around $\Delta \Omega = 0^\circ$. Runs where the probability of detecting at least 3 transiting planets is greater than the probability of detecting two transiting planets (top) show nodal alignment, while the other runs (bottom) do not.}
  \label{fig:polar}
\end{figure*}

Some of these results can be gleaned from Figure \ref{fig:polar}. The figure shows the inclination and ascending node of every planet inside 200 days. We exclude simulations with fewer than three planets inside 200 days because it does not make sense to talk about nodal alignment for a two-planet system. The 200-day cut-off removes most simulations from model B, but it is important to focus our analysis to the group of planets that are actually inside Kepler's detection window. The inclination is computed relative to the invariant plane. For the ascending node we use $\Delta \Omega = \Omega - \overline{\Omega}$ where $\overline{\Omega}$ is the average of all ascending nodes within one system. In other words, if the ascending nodes are aligned, $\Delta \Omega$ values will be clustered near the $\Delta \Omega = 0^\circ$ line.

All plots in Figure \ref{fig:polar} show show some degree of clustering toward $\Delta \Omega = 0^\circ$, indicating nodal alignment. We find that, as expected, systems where the probability of detecting three or more transiting planets (``$\rm Pr(3+t)$'') is greater than the probability of detecting exactly two transiting planets (``$\rm Pr(2t)$'') show a higher degree of nodal alignment than the other systems.

Finally, we conducted one final thought experiment: We randomised the ascending nodes of all the planets in both models and performed a new set of simulated Kepler observations. Eliminating nodal alignment slightly increased the percentage of 1-transit systems by about 1.5 percentage points in both models, and slightly increased the percentage of 2-transit systems by 0.9 percentage points in both models (see Table \ref{tab:nodal_alignment}). However, the fractional change in the number of 3+ transiting systems was more significant; eliminating nodal alignment actually decreased the percentage of 3+ transiting systems by 2.5\% and 0.5\% for models A and B, respectively. If planetary systems indeed form with some amount of nodal alignment, as in our simulations, then that alignment will increase the frequency of multiple planet systems.

\begin{table}
  \centering
  \caption{Effect of nodal alignment in transit multiplicities. The second and fourth rows are the result of randomising the arguments of ascending nodes before performing simulated Kepler observations. The columns labelled ``1t'', ``2t'', and ``3+t'' show the fraction of observations that reveal one, two, or more transiting planets. Figures for the Kepler sample are also included for comparison.}
  \label{tab:nodal_alignment}
  \begin{tabular}{llrrr}
  Sample & Randomized nodes & 1t$\;\;\;\;$ & 2t$\;\;\;\;$ & 3+t$\;\;\;$\\
  \hline
  Model A     & No & 65.4\% & 22.1\% & 12.5\% \\
  Model A     & Yes  & 67.0\% & 23.0\% & 10.0\% \\
  \hline
  Model B     & No & 90.9\% &  5.0\% &  4.1\% \\
  Model B     & Yes  & 92.3\% &  4.1\% &  3.6\% \\
  \hline
  Kepler & N/A & 80.47\% & 14.12\% & 5.41\% \\
  \end{tabular}
\end{table}
%
%
\section{Discussion}
\label{sec:discussion}

\subsection{Implications for statistical models}
\label{sec:discussion:stats}

Most statistical models of planetary systems treat orbital parameters and planet masses as largely independent variables (e.g. \citet{He_2019} draws the parameters independently and rejects those that are likely to be unstable). Our work shows that during the formation process, these parameters may become correlated in ways that impact key observables like the number of transiting planets and their period ratios.

\begin{itemize}

\item First, much of the correlation between orbital separations and eccentricity is probably well captured by a simple stability test that requires that orbits do not cross \citep[e.g.][]{Hsu_2019} or forces a minimum separation between the apastron of one orbit and the periastron of the next \citep[e.g.][]{He_2019}
\begin{equation}
a_2(1 - e_2) - a_1(1 + e_1) > \Delta_{\rm crit} / R_{\rm Hill}
\label{eqn:HillStabilityApoPeri}
\end{equation}
\item The correlation between eccentricity and mutual inclination requires a more sophisticated notion of dynamical stability. A good strategy adopted by \citet{He_2020} is to determine the maximum amount of angular momentum deficit (AMD) that allows for secular stability \citep{Laskar_2017}, and assume some form of AMD equipartition of energy between epicyclic motion in the vertical and radial directions that sets the orbital inclinations and eccentricities. \citet{He_2020} have shown that this strategy naturally reproduces many observed features of the Kepler sample.

\item We are not aware of any statistical model that captures the correlation between planet masses and disc migration (i.e. high-mass systems experience more disc migration). In fact, the simple stability criteria in Eqn. \ref{eqn:HillStabilityApoPeri} would be expected to produce a correlation with the opposite sign since it is easier for a naive simulation (i.e., one without migration or N-body dynamics) to produce a compact stable system from low-mass planets. We suggest that future research further investigate this correlation by introducing a new parameter, e.g., the critical mass of the co-rotation trap.

\item Finally, statistical models usually assign the planets uniform random longitudes of ascending node. We have shown that ascending nodes are not distributed uniformly and that randomising them will slightly increase the frequency of 1-transit systems relative to multiple transiting systems. However, the effect is somewhat small, increasing the frequency of 1-transit systems by only about 1.5\%, increasing the frequency of 2-transit systems by only about 1.5\%, and decreasing the frequency of 3+ transiting systems by 0.5-2.5\%. It is unclear whether this effect is small enough to avoid significantly impacting statistical studies.
\end{itemize}

\subsection{Implications for disk structure models}
\label{sec:discussion:disks}

Our results point to a deep connection between Kepler's systems with tightly packed inner planets (STIPs) and the thermal structure of the parent protoplanetary discs, as it sets the balance between co-rotation and Lindblad torques. However, this study only scratches the surface of this connection. Ideally, future work would help answer the question of whether the masses of observed STIPs can be used to constrain the structure of their parent discs. To address this question, we recommend future investigations run simulations where the disc thermal gradient is a simple powerlaw and investigate how the exponent affects the frequency and mass of STIPs planets.

\subsection{Planet clusters and stability}
\label{sec:discussion:clusters}

The stabilizing effect of separating the planets into distinct clusters does not seem to be captured by commonly used stability criteria, or by AMD stability \citep{Laskar_2017}. Case in point, the solar system has survived for billions of years despite being ``AMD unstable''. The reason for that longevity is that the terrestrial planets are relatively poorly coupled to the outer planets. We can see two concrete examples of this:

\textit{First}, the inner terrestrial planets would be an AMD stable system on their own. The reason that the solar system is AMD unstable is because the giant planets hold so much AMD that if it were transferred to the terrestrial planets, the terrestrial planets could easily cross orbits.

\textit{Second}, it is relatively easy to show that leading-order secular interactions alone will never trigger close encounters \citep[but higher order terms can lead to instability, as shown by][]{Batygin_2015}. To see this, let $(h_j, k_j) = (e_j \sin\varpi_j, e_j \cos\varpi_j)$ be the complex eccentricity of the $j^{\rm th}$ planet, where $e_j$ is the planet's eccentricity and $\varpi_j$ is its longitude of pericenter. A full review of secular theory is beyond the scope of this work \citep[for that, see][]{Murray_Dermott_1999}, but briefly, the orbital evolution of a co-planar system of $N$ spherical planets can be written as
\begin{eqnarray}
h_j &=& \sum_i^N e_{ji} \sin(g_i t + \beta_i) \\
k_j &=& \sum_i^N e_{ji} \cos(g_i t + \beta_i)
\end{eqnarray}
where the terms $e_{ji}$, $g_i$, and $\beta_i$ are set by the initial conditions $(a_j, e_j, \varpi_j)$. The precise definition of these terms can be found in chapter 6 of \citet{Murray_Dermott_1999}. Computing them is slightly tedious but not difficult (around 160 lines of code). With this we can easily compute an upper bound on each planet's eccentricity
\begin{equation}\label{eqn:e_jmax}
e_j^2 \equiv h_j^2 + k_j^2 
      \le e_{j,\rm max}^2 \equiv \sum_i^N \sum_k^N |e_{ji} e_{jk}|
\end{equation}

When we compute $e_{j,\rm max}$ for the solar system we find that none of the solar system planets can ever cross orbit. Of course, higher order terms make the true long-term behaviour of the solar system far more complex \citep{Batygin_2015}. But the key point is that a quick calculation of $e_{j,\rm max}$ as defined in Equation \ref{eqn:e_jmax} can identify systems with the potential to be long-lived.

Perhaps the most important limitation of the AMD stability criterion is that it allows unconstrained transfer of angular momentum. While AMD instability is one of the best tools available today, we caution that planetary systems with well-separated clusters of planets may be effectively stable for gigayear timescales.  Further research in how to apply secular theory could prove valuable for identifying AMD unstable systems that may nevertheless be long-lived.

%
%
\section{Conclusion}
\label{sec:conclusion}

Numerous studies have attempted to interpret the relative rate of planetary systems observed to have one or more transiting planets by NASA's Kepler mission \citep[e.g.][]{Lissauer_2011,Fabrycky_2014}. Several authors have proposed that Kepler systems are composed of two distinct exoplanet populations: a dynamically hot population characterised by high mutual inclinations, and a dynamically cold population \citep{Johansen_2012,Moriarty_2016,Izidoro_2017}.  
In this paper we present sets of planet formation simulations that share many of the key properties of Kepler's planetary systems. We probe the formation histories of our planetary systems and investigate the architectural properties that cause systems to be identified as a single-transiting planet system. We show that a bimodality in inclinations is neither required by observation, nor is it supported by theory (Figure \ref{fig:dichotomy-hist}). Instead, the differences in mutual orbital inclinations between planetary systems with one transiting planet and those with multiple transiting planets are dynamically insignificant.

We find that the number of transiting planets does not stem from qualitative differences in the formation history of the planetary systems. However, each of our models has a large spike of planetary systems with a very high probability of being detected as single transit (Figure \ref{fig:thought}). Since the distribution of mutual inclinations is similar, the strongest predictor of low transit multiplicities is a planetary system that only has a small number of detectable planets in the first place (see \S~\ref{sec:results:dichotomy}). This implies that Kepler's multiple planetary systems could be much more representative of planetary systems in general than previously thought.

We propose an alternative explanation for the apparent Kepler dichotomy. We suggest that orbital migration traps break planetary system into clusters, causing some systems to have only a small number of Kepler-detectable planets (see \S~\ref{sec:results:clusters}). We found that some simulations build relatively large (super-Earth size) planets whose migration is completely dominated by Lindblad torques. These planets migrate rapidly and form a cluster of short period planets that is relatively isolated from the dynamical instabilities that typically occur in the outer system (Figure \ref{fig:snapshots}, left). Other simulations produce more planets which are smaller (roughly Earth-sized) and become caught in the co-rotation trap, causing them to retain larger periods and remain more dynamically coupled to the outer planets. The first formation history is associated with more planets inside the Kepler detection region, for two reasons:

\begin{itemize}
\item More migration and shorter periods make it easier to fit more planets into the inner region of the planetary system (which this study defines as orbital periods less than $\sim$200 days).

\item A population of planetary systems with fewer late-stage instabilities is more likely to lead to compact planetary systems. In turn, this makes it easier to fit more planets in the inner region of the planetary system.
\end{itemize}

\noindent We find that the first formation history is frequently associated with simulations that contain a higher total solid mass (i.e. model A). The model B simulations contained half the total solid mass of model A, and we do not observe this strong clustering effect.

In addition, instabilities that produce larger separations and fewer planets in the inner planetary system also cause high mutual inclinations, which can further decrease the probability that more than one planet will transit the same line of sight. Finally, the smaller planets that get caught in the co-rotation trap also have smaller transit depths and thus lower detection probability.  Kepler's lower detection efficiency for smaller planets means that even if multiple planets transit, there is a disproportionately lower probability that multiple planets will be discovered.

One of the most interesting aspects of these results is that all of these variables could naturally become correlated in the planet formation process.  Systems with shorter orbital periods, are more likely to: (1) be more compact, (2) host more planets inside 200 days, (3) have lower mutual inclinations, and (4) have larger transit depths. Perhaps the most interesting aspect is that all of this is connected to the co-rotation torque and therefore the structure of the protoplanetary disc. In other words, if our proposed explanation for the apparent Kepler dichotomy is correct, then sizes of Kepler's systems of tightly packed inner planets (STIPs) may be a probe into the otherwise unobservable structure of the inner protoplanetary discs where they formed.
This prediction can be tested via extremely precise radial velocity follow-up observations, targeting planets with single and multiple transiting planets.

%
%
\begin{acknowledgments}
We acknowledge support from NASA Exoplanet Research Program award NNX15AE21G. The results reported herein benefited from collaborations and/or information exchange within NASA’s Nexus for Exoplanet System Science (NExSS) research coordination network sponsored by NASA’s Science Mission Directorate. The Center for Exoplanets and Habitable Worlds is supported by the Pennsylvania State University, the Eberly College of Science, and the Pennsylvania Space Grant Consortium. We gratefully acknowledge support from National Science Foundation (NSF) grant MRI-1626251. 
This work was supported by a grant from the Simons Foundation/SFARI (675601, E.B.F.).
E.B.F. acknowledges the support of the Ambrose Monell Foundation and the Institute for Advanced Study.  
This research or portions of this research were conducted with Advanced Cyber Infrastructure computational resources provided by The Institute for Computational \& Data Sciences at The Pennsylvania State University (http://icds.psu.edu), including the CyberLAMP cluster supported by NSF grant MRI-1626251. This research has made use of the NASA Exoplanet Archive \citep[][accessed on 2021-08-10 at 08:26 and returning 2121 rows]{koidr25}, which is operated by the California Institute of Technology, under contract with the National Aeronautics and Space Administration under the Exoplanet Exploration Program.
\end{acknowledgments}


\bibliography{refs}{}

\begin{thebibliography}{}
\expandafter\ifx\csname natexlab\endcsname\relax\def\natexlab#1{#1}\fi
\providecommand{\url}[1]{\href{#1}{#1}}
\providecommand{\dodoi}[1]{doi:~\href{http://doi.org/#1}{\nolinkurl{#1}}}
\providecommand{\doeprint}[1]{\href{http://ascl.net/#1}{\nolinkurl{http://ascl.net/#1}}}
\providecommand{\doarXiv}[1]{\href{https://arxiv.org/abs/#1}{\nolinkurl{https://arxiv.org/abs/#1}}}

\bibitem[{{Armitage}(2007)}]{Armitage_2007}
{Armitage}, P.~J. 2007, arXiv Astrophysics e-prints

\bibitem[{{Bai} \& {Stone}(2010)}]{Bai_2010}
{Bai}, X.-N., \& {Stone}, J.~M. 2010, \apj, 722, 1437,
  \dodoi{10.1088/0004-637X/722/2/1437}

\bibitem[{{Ballard} \& {Johnson}(2016)}]{Ballard_2016}
{Ballard}, S., \& {Johnson}, J.~A. 2016, \apj, 816, 66,
  \dodoi{10.3847/0004-637X/816/2/66}

\bibitem[{{Batalha} {et~al.}(2013){Batalha}, {Rowe}, {Bryson}, {Barclay},
  {Burke}, {Caldwell}, {Christiansen}, {Mullally}, {Thompson}, {Brown},
  {Dupree}, {Fabrycky}, {Ford}, {Fortney}, {Gilliland}, {Isaacson}, {Latham},
  {Marcy}, {Quinn}, {Ragozzine}, {Shporer}, {Borucki}, {Ciardi}, {Gautier},
  {Haas}, {Jenkins}, {Koch}, {Lissauer}, {Rapin}, {Basri}, {Boss}, {Buchhave},
  {Carter}, {Charbonneau}, {Christensen-Dalsgaard}, {Clarke}, {Cochran},
  {Demory}, {Desert}, {Devore}, {Doyle}, {Esquerdo}, {Everett}, {Fressin},
  {Geary}, {Girouard}, {Gould}, {Hall}, {Holman}, {Howard}, {Howell},
  {Ibrahim}, {Kinemuchi}, {Kjeldsen}, {Klaus}, {Li}, {Lucas}, {Meibom},
  {Morris}, {Pr{\v s}a}, {Quintana}, {Sanderfer}, {Sasselov}, {Seader},
  {Smith}, {Steffen}, {Still}, {Stumpe}, {Tarter}, {Tenenbaum}, {Torres},
  {Twicken}, {Uddin}, {Van Cleve}, {Walkowicz}, \& {Welsh}}]{Batalha_2013}
{Batalha}, N.~M., {Rowe}, J.~F., {Bryson}, S.~T., {et~al.} 2013, \apjs, 204,
  24, \dodoi{10.1088/0067-0049/204/2/24}

\bibitem[{{Batygin} \& {Laughlin}(2008)}]{Batygin_2008}
{Batygin}, K., \& {Laughlin}, G. 2008, \apj, 683, 1207, \dodoi{10.1086/589232}

\bibitem[{{Batygin} {et~al.}(2015){Batygin}, {Morbidelli}, \&
  {Holman}}]{Batygin_2015}
{Batygin}, K., {Morbidelli}, A., \& {Holman}, M.~J. 2015, \apj, 799, 120,
  \dodoi{10.1088/0004-637X/799/2/120}

\bibitem[{{Birnstiel} {et~al.}(2012){Birnstiel}, {Klahr}, \&
  {Ercolano}}]{Birnstiel_2012}
{Birnstiel}, T., {Klahr}, H., \& {Ercolano}, B. 2012, \aap, 539, A148,
  \dodoi{10.1051/0004-6361/201118136}

\bibitem[{{Bitsch} {et~al.}(2015){Bitsch}, {Johansen}, {Lambrechts}, \&
  {Morbidelli}}]{Bitsch_2015}
{Bitsch}, B., {Johansen}, A., {Lambrechts}, M., \& {Morbidelli}, A. 2015, \aap,
  575, A28, \dodoi{10.1051/0004-6361/201424964}

\bibitem[{{Blum}(2018)}]{Blum_2018}
{Blum}, J. 2018, \ssr, 214, 52, \dodoi{10.1007/s11214-018-0486-5}

\bibitem[{{Borucki} {et~al.}(2010){Borucki}, {Koch}, {Basri}, {Batalha},
  {Brown}, {Caldwell}, {Caldwell}, {Christensen-Dalsgaard}, {Cochran},
  {DeVore}, {Dunham}, {Dupree}, {Gautier}, {Geary}, {Gilliland}, {Gould},
  {Howell}, {Jenkins}, {Kondo}, {Latham}, {Marcy}, {Meibom}, {Kjeldsen},
  {Lissauer}, {Monet}, {Morrison}, {Sasselov}, {Tarter}, {Boss}, {Brownlee},
  {Owen}, {Buzasi}, {Charbonneau}, {Doyle}, {Fortney}, {Ford}, {Holman},
  {Seager}, {Steffen}, {Welsh}, {Rowe}, {Anderson}, {Buchhave}, {Ciardi},
  {Walkowicz}, {Sherry}, {Horch}, {Isaacson}, {Everett}, {Fischer}, {Torres},
  {Johnson}, {Endl}, {MacQueen}, {Bryson}, {Dotson}, {Haas}, {Kolodziejczak},
  {Van Cleve}, {Chandrasekaran}, {Twicken}, {Quintana}, {Clarke}, {Allen},
  {Li}, {Wu}, {Tenenbaum}, {Verner}, {Bruhweiler}, {Barnes}, \&
  {Prsa}}]{Borucki_2010}
{Borucki}, W.~J., {Koch}, D., {Basri}, G., {et~al.} 2010, Science, 327, 977,
  \dodoi{10.1126/science.1185402}

\bibitem[{{Bovaird} \& {Lineweaver}(2017)}]{Bovaird_2017}
{Bovaird}, T., \& {Lineweaver}, C.~H. 2017, \mnras, 468, 1493,
  \dodoi{10.1093/mnras/stx414}

\bibitem[{{Brasser} {et~al.}(2018){Brasser}, {Matsumura}, {Muto}, \&
  {Ida}}]{Brasser_2018}
{Brasser}, R., {Matsumura}, S., {Muto}, T., \& {Ida}, S. 2018, \apjl, 864, L8,
  \dodoi{10.3847/2041-8213/aada18}

\bibitem[{{Carrera} {et~al.}(2019){Carrera}, {Ford}, \&
  {Izidoro}}]{Carrera_2019}
{Carrera}, D., {Ford}, E.~B., \& {Izidoro}, A. 2019, \mnras, 963,
  \dodoi{10.1093/mnras/stz974}

\bibitem[{{Carrera} {et~al.}(2015){Carrera}, {Johansen}, \&
  {Davies}}]{Carrera_2015}
{Carrera}, D., {Johansen}, A., \& {Davies}, M.~B. 2015, \aap, 579, A43,
  \dodoi{10.1051/0004-6361/201425120}

\bibitem[{{Chambers}(2006)}]{Chambers_2006}
{Chambers}, J. 2006, \icarus, 180, 496, \dodoi{10.1016/j.icarus.2005.10.017}

\bibitem[{{Chambers}(1999)}]{Chambers_1999}
{Chambers}, J.~E. 1999, \mnras, 304, 793,
  \dodoi{10.1046/j.1365-8711.1999.02379.x}

\bibitem[{{Chambers} {et~al.}(1996){Chambers}, {Wetherill}, \&
  {Boss}}]{Chambers_1996}
{Chambers}, J.~E., {Wetherill}, G.~W., \& {Boss}, A.~P. 1996, \icarus, 119,
  261, \dodoi{10.1006/icar.1996.0019}

\bibitem[{{Chatterjee} {et~al.}(2008){Chatterjee}, {Ford}, {Matsumura}, \&
  {Rasio}}]{Chatterjee_2008}
{Chatterjee}, S., {Ford}, E.~B., {Matsumura}, S., \& {Rasio}, F.~A. 2008, \apj,
  686, 580, \dodoi{10.1086/590227}

\bibitem[{{Coleman} \& {Nelson}(2014)}]{Coleman_2014}
{Coleman}, G.~A.~L., \& {Nelson}, R.~P. 2014, \mnras, 445, 479,
  \dodoi{10.1093/mnras/stu1715}

\bibitem[{{Cresswell} \& {Nelson}(2006)}]{Cresswell_2006}
{Cresswell}, P., \& {Nelson}, R.~P. 2006, \aap, 450, 833,
  \dodoi{10.1051/0004-6361:20054551}

\bibitem[{{Cresswell} \& {Nelson}(2008)}]{Cresswell_2008}
---. 2008, \aap, 482, 677, \dodoi{10.1051/0004-6361:20079178}

\bibitem[{{Dawson} {et~al.}(2016){Dawson}, {Lee}, \& {Chiang}}]{Dawson_2016}
{Dawson}, R.~I., {Lee}, E.~J., \& {Chiang}, E. 2016, \apj, 822, 54,
  \dodoi{10.3847/0004-637X/822/1/54}

\bibitem[{{Faber} \& {Quillen}(2007)}]{Faber_2007}
{Faber}, P., \& {Quillen}, A.~C. 2007, \mnras, 382, 1823,
  \dodoi{10.1111/j.1365-2966.2007.12490.x}

\bibitem[{{Fabrycky} {et~al.}(2014){Fabrycky}, {Lissauer}, {Ragozzine}, {Rowe},
  {Steffen}, {Agol}, {Barclay}, {Batalha}, {Borucki}, {Ciardi}, {Ford},
  {Gautier}, {Geary}, {Holman}, {Jenkins}, {Li}, {Morehead}, {Morris},
  {Shporer}, {Smith}, {Still}, \& {Van Cleve}}]{Fabrycky_2014}
{Fabrycky}, D.~C., {Lissauer}, J.~J., {Ragozzine}, D., {et~al.} 2014, \apj,
  790, 146, \dodoi{10.1088/0004-637X/790/2/146}

\bibitem[{{Fendyke} \& {Nelson}(2014)}]{Fendyke_2014}
{Fendyke}, S.~M., \& {Nelson}, R.~P. 2014, \mnras, 437, 96,
  \dodoi{10.1093/mnras/stt1867}

\bibitem[{{Goldreich} \& {Tremaine}(1979)}]{Goldreich_1979}
{Goldreich}, P., \& {Tremaine}, S. 1979, \apj, 233, 857, \dodoi{10.1086/157448}

\bibitem[{{Greenberg} {et~al.}(1978){Greenberg}, {Wacker}, {Hartmann}, \&
  {Chapman}}]{Greenberg_1978}
{Greenberg}, R., {Wacker}, J.~F., {Hartmann}, W.~K., \& {Chapman}, C.~R. 1978,
  \icarus, 35, 1, \dodoi{10.1016/0019-1035(78)90057-X}

\bibitem[{{G{\"u}ttler} {et~al.}(2010){G{\"u}ttler}, {Blum}, {Zsom}, {Ormel},
  \& {Dullemond}}]{Guettler_2010}
{G{\"u}ttler}, C., {Blum}, J., {Zsom}, A., {Ormel}, C.~W., \& {Dullemond},
  C.~P. 2010, \aap, 513, A56, \dodoi{10.1051/0004-6361/200912852}

\bibitem[{{He} {et~al.}(2019){He}, {Ford}, \& {Ragozzine}}]{He_2019}
{He}, M.~Y., {Ford}, E.~B., \& {Ragozzine}, D. 2019, \mnras, 490, 4575,
  \dodoi{10.1093/mnras/stz2869}

\bibitem[{{He} {et~al.}(2020){He}, {Ford}, {Ragozzine}, \& {Carrera}}]{He_2020}
{He}, M.~Y., {Ford}, E.~B., {Ragozzine}, D., \& {Carrera}, D. 2020, \aj, 160,
  276, \dodoi{10.3847/1538-3881/abba18}

\bibitem[{{Hsu} {et~al.}(2019){Hsu}, {Ford}, {Ragozzine}, \&
  {Ashby}}]{Hsu_2019}
{Hsu}, D.~C., {Ford}, E.~B., {Ragozzine}, D., \& {Ashby}, K. 2019, \aj, 158,
  109, \dodoi{10.3847/1538-3881/ab31ab}

\bibitem[{{Inamdar} \& {Schlichting}(2015)}]{Inamdar_2015}
{Inamdar}, N.~K., \& {Schlichting}, H.~E. 2015, \mnras, 448, 1751,
  \dodoi{10.1093/mnras/stv030}

\bibitem[{{Izidoro} {et~al.}(2021){Izidoro}, {Bitsch}, {Raymond}, {Johansen},
  {Morbidelli}, {Lambrechts}, \& {Jacobson}}]{Izidoro_2021}
{Izidoro}, A., {Bitsch}, B., {Raymond}, S.~N., {et~al.} 2021, \aap, 650, A152,
  \dodoi{10.1051/0004-6361/201935336}

\bibitem[{{Izidoro} {et~al.}(2017){Izidoro}, {Ogihara}, {Raymond},
  {Morbidelli}, {Pierens}, {Bitsch}, {Cossou}, \& {Hersant}}]{Izidoro_2017}
{Izidoro}, A., {Ogihara}, M., {Raymond}, S.~N., {et~al.} 2017, \mnras, 470,
  1750, \dodoi{10.1093/mnras/stx1232}

\bibitem[{{Johansen} {et~al.}(2012){Johansen}, {Davies}, {Church}, \&
  {Holmelin}}]{Johansen_2012}
{Johansen}, A., {Davies}, M.~B., {Church}, R.~P., \& {Holmelin}, V. 2012, \apj,
  758, 39, \dodoi{10.1088/0004-637X/758/1/39}

\bibitem[{{Johansen} {et~al.}(2007){Johansen}, {Oishi}, {Mac Low}, {Klahr},
  {Henning}, \& {Youdin}}]{Johansen_2007}
{Johansen}, A., {Oishi}, J.~S., {Mac Low}, M.-M., {et~al.} 2007, \nat, 448,
  1022, \dodoi{10.1038/nature06086}

\bibitem[{{Juri{\'c}} \& {Tremaine}(2008)}]{Juric_2008}
{Juri{\'c}}, M., \& {Tremaine}, S. 2008, \apj, 686, 603, \dodoi{10.1086/590047}

\bibitem[{{Kley} \& {Nelson}(2012)}]{Kley_2012}
{Kley}, W., \& {Nelson}, R.~P. 2012, \araa, 50, 211,
  \dodoi{10.1146/annurev-astro-081811-125523}

\bibitem[{{Kokubo} \& {Ida}(1996)}]{Kokubo_1996}
{Kokubo}, E., \& {Ida}, S. 1996, \icarus, 123, 180,
  \dodoi{10.1006/icar.1996.0148}

\bibitem[{{Kokubo} \& {Ida}(2000)}]{Kokubo_2000}
---. 2000, \icarus, 143, 15, \dodoi{10.1006/icar.1999.6237}

\bibitem[{{Lai} \& {Pu}(2017)}]{Dong_2017}
{Lai}, D., \& {Pu}, B. 2017, \aj, 153, 42, \dodoi{10.3847/1538-3881/153/1/42}

\bibitem[{{Lambrechts} \& {Johansen}(2012)}]{Lambrechts_2012}
{Lambrechts}, M., \& {Johansen}, A. 2012, \aap, 544, A32,
  \dodoi{10.1051/0004-6361/201219127}

\bibitem[{{Laskar} \& {Petit}(2017)}]{Laskar_2017}
{Laskar}, J., \& {Petit}, A.~C. 2017, \aap, 605, A72,
  \dodoi{10.1051/0004-6361/201630022}

\bibitem[{{Levison} {et~al.}(2015){Levison}, {Kretke}, \&
  {Duncan}}]{Levison_2015}
{Levison}, H.~F., {Kretke}, K.~A., \& {Duncan}, M.~J. 2015, \nat, 524, 322,
  \dodoi{10.1038/nature14675}

\bibitem[{{Lissauer} {et~al.}(2011){Lissauer}, {Ragozzine}, {Fabrycky},
  {Steffen}, {Ford}, {Jenkins}, {Shporer}, {Holman}, {Rowe}, {Quintana},
  {Batalha}, {Borucki}, {Bryson}, {Caldwell}, {Carter}, {Ciardi}, {Dunham},
  {Fortney}, {Gautier}, {Howell}, {Koch}, {Latham}, {Marcy}, {Morehead}, \&
  {Sasselov}}]{Lissauer_2011}
{Lissauer}, J.~J., {Ragozzine}, D., {Fabrycky}, D.~C., {et~al.} 2011, \apjs,
  197, 8, \dodoi{10.1088/0067-0049/197/1/8}

\bibitem[{{MacDonald} {et~al.}(2020){MacDonald}, {Dawson}, {Morrison}, {Lee},
  \& {Khandelwal}}]{MacDonald_2020}
{MacDonald}, M.~G., {Dawson}, R.~I., {Morrison}, S.~J., {Lee}, E.~J., \&
  {Khandelwal}, A. 2020, \apj, 891, 20, \dodoi{10.3847/1538-4357/ab6f04}

\bibitem[{{Millholland} {et~al.}(2021){Millholland}, {He}, {Ford}, {Ragozzine},
  {Fabrycky}, \& {Winn}}]{Millholland_2021}
{Millholland}, S.~C., {He}, M.~Y., {Ford}, E.~B., {et~al.} 2021, \aj, 162, 166,
  \dodoi{10.3847/1538-3881/ac0f7a}

\bibitem[{{Millholland} {et~al.}(2022){Millholland}, {He}, \&
  {Zink}}]{Millholland_2022}
{Millholland}, S.~C., {He}, M.~Y., \& {Zink}, J.~K. 2022, arXiv e-prints,
  arXiv:2207.10068.
\newblock \doarXiv{2207.10068}

\bibitem[{{Moriarty} \& {Ballard}(2016)}]{Moriarty_2016}
{Moriarty}, J., \& {Ballard}, S. 2016, \apj, 832, 34,
  \dodoi{10.3847/0004-637X/832/1/34}

\bibitem[{{Mulders} {et~al.}(2019){Mulders}, {Mordasini}, {Pascucci}, {Ciesla},
  {Emsenhuber}, \& {Apai}}]{Mulders_2019}
{Mulders}, G.~D., {Mordasini}, C., {Pascucci}, I., {et~al.} 2019, \apj, 887,
  157, \dodoi{10.3847/1538-4357/ab5187}

\bibitem[{{Mulders} {et~al.}(2020){Mulders}, {O'Brien}, {Ciesla}, {Apai}, \&
  {Pascucci}}]{Mulders_2020}
{Mulders}, G.~D., {O'Brien}, D.~P., {Ciesla}, F.~J., {Apai}, D., \& {Pascucci},
  I. 2020, \apj, 897, 72, \dodoi{10.3847/1538-4357/ab9806}

\bibitem[{{Mulders} {et~al.}(2018){Mulders}, {Pascucci}, {Apai}, \&
  {Ciesla}}]{Mulders_2018}
{Mulders}, G.~D., {Pascucci}, I., {Apai}, D., \& {Ciesla}, F.~J. 2018, \aj,
  156, 24, \dodoi{10.3847/1538-3881/aac5ea}

\bibitem[{{Murray} \& {Dermott}(1999)}]{Murray_Dermott_1999}
{Murray}, C.~D., \& {Dermott}, S.~F. 1999, {Solar system dynamics}

\bibitem[{{NASA Exoplanet Archive}(2021)}]{koidr25}
{NASA Exoplanet Archive}. 2021, Kepler Objects of Interest DR25, Version:
  2021-08-10 08:26,  NExScI-Caltech/IPAC, \dodoi{10.26133/NEA5}

\bibitem[{{Nesvorn{\'y}} {et~al.}(2019){Nesvorn{\'y}}, {Li}, {Youdin}, {Simon},
  \& {Grundy}}]{Nesvorny_2019}
{Nesvorn{\'y}}, D., {Li}, R., {Youdin}, A.~N., {Simon}, J.~B., \& {Grundy},
  W.~M. 2019, Nature Astronomy, 3, 808, \dodoi{10.1038/s41550-019-0806-z}

\bibitem[{{Paardekooper} {et~al.}(2010){Paardekooper}, {Baruteau}, {Crida}, \&
  {Kley}}]{Paardekooper_2010}
{Paardekooper}, S.-J., {Baruteau}, C., {Crida}, A., \& {Kley}, W. 2010, \mnras,
  401, 1950, \dodoi{10.1111/j.1365-2966.2009.15782.x}

\bibitem[{{Paardekooper} {et~al.}(2011){Paardekooper}, {Baruteau}, \&
  {Kley}}]{Paardekooper_2011}
{Paardekooper}, S.-J., {Baruteau}, C., \& {Kley}, W. 2011, \mnras, 410, 293,
  \dodoi{10.1111/j.1365-2966.2010.17442.x}

\bibitem[{{Papaloizou} \& {Larwood}(2000)}]{Papaloizou_2000}
{Papaloizou}, J.~C.~B., \& {Larwood}, J.~D. 2000, \mnras, 315, 823,
  \dodoi{10.1046/j.1365-8711.2000.03466.x}

\bibitem[{{Petit} {et~al.}(2017){Petit}, {Laskar}, \& {Bou{\'e}}}]{Petit_2017}
{Petit}, A.~C., {Laskar}, J., \& {Bou{\'e}}, G. 2017, \aap, 607, A35,
  \dodoi{10.1051/0004-6361/201731196}

\bibitem[{{Pu} \& {Wu}(2015)}]{Pu_2015}
{Pu}, B., \& {Wu}, Y. 2015, \apj, 807, 44, \dodoi{10.1088/0004-637X/807/1/44}

\bibitem[{{Rasio} \& {Ford}(1996)}]{Rasio_1996}
{Rasio}, F.~A., \& {Ford}, E.~B. 1996, Science, 274, 954,
  \dodoi{10.1126/science.274.5289.954}

\bibitem[{{Shakura} \& {Sunyaev}(1973)}]{Shakura_1973}
{Shakura}, N.~I., \& {Sunyaev}, R.~A. 1973, \aap, 24, 337

\bibitem[{{Spalding} \& {Batygin}(2016)}]{Spalding_2016}
{Spalding}, C., \& {Batygin}, K. 2016, \apj, 830, 5,
  \dodoi{10.3847/0004-637X/830/1/5}

\bibitem[{{Tanaka} \& {Ward}(2004)}]{Tanaka_2004}
{Tanaka}, H., \& {Ward}, W.~R. 2004, \apj, 602, 388, \dodoi{10.1086/380992}

\bibitem[{{Terquem} \& {Papaloizou}(2007)}]{Terquem_2007}
{Terquem}, C., \& {Papaloizou}, J.~C.~B. 2007, \apj, 654, 1110,
  \dodoi{10.1086/509497}

\bibitem[{{Thommes} {et~al.}(2003){Thommes}, {Duncan}, \&
  {Levison}}]{Thommes_2003}
{Thommes}, E.~W., {Duncan}, M.~J., \& {Levison}, H.~F. 2003, \icarus, 161, 431,
  \dodoi{10.1016/S0019-1035(02)00043-X}

\bibitem[{{Van Eylen} {et~al.}(2019){Van Eylen}, {Albrecht}, {Huang},
  {MacDonald}, {Dawson}, {Cai}, {Foreman-Mackey}, {Lundkvist}, {Silva Aguirre},
  {Snellen}, \& {Winn}}]{VanEylen_2019}
{Van Eylen}, V., {Albrecht}, S., {Huang}, X., {et~al.} 2019, \aj, 157, 61,
  \dodoi{10.3847/1538-3881/aaf22f}

\bibitem[{{Volk} \& {Gladman}(2015)}]{Volk_2015}
{Volk}, K., \& {Gladman}, B. 2015, \apjl, 806, L26,
  \dodoi{10.1088/2041-8205/806/2/L26}

\bibitem[{{Weidenschilling}(1977)}]{Weidenschilling_1977}
{Weidenschilling}, S.~J. 1977, \mnras, 180, 57, \dodoi{10.1093/mnras/180.1.57}

\bibitem[{{Wetherill} \& {Stewart}(1989)}]{Wetherill_1989}
{Wetherill}, G.~W., \& {Stewart}, G.~R. 1989, \icarus, 77, 330,
  \dodoi{10.1016/0019-1035(89)90093-6}

\bibitem[{{Yang} {et~al.}(2017){Yang}, {Johansen}, \& {Carrera}}]{Yang_2017}
{Yang}, C.~C., {Johansen}, A., \& {Carrera}, D. 2017, \aap, 606, A80,
  \dodoi{10.1051/0004-6361/201630106}

\bibitem[{{Youdin} \& {Goodman}(2005)}]{Youdin_2005}
{Youdin}, A.~N., \& {Goodman}, J. 2005, \apj, 620, 459, \dodoi{10.1086/426895}

\bibitem[{{Zeng} {et~al.}(2016){Zeng}, {Sasselov}, \& {Jacobsen}}]{Zeng_2016}
{Zeng}, L., {Sasselov}, D.~D., \& {Jacobsen}, S.~B. 2016, \apj, 819, 127,
  \dodoi{10.3847/0004-637X/819/2/127}

\bibitem[{{Zink} {et~al.}(2019){Zink}, {Christiansen}, \& {Hansen}}]{Zink_2019}
{Zink}, J.~K., {Christiansen}, J.~L., \& {Hansen}, B. M.~S. 2019, \mnras, 483,
  4479, \dodoi{10.1093/mnras/sty3463}

\bibitem[{{Zsom} {et~al.}(2010){Zsom}, {Ormel}, {G{\"u}ttler}, {Blum}, \&
  {Dullemond}}]{Zsom_2010}
{Zsom}, A., {Ormel}, C.~W., {G{\"u}ttler}, C., {Blum}, J., \& {Dullemond},
  C.~P. 2010, \aap, 513, A57, \dodoi{10.1051/0004-6361/200912976}

\end{thebibliography}
\bibliographystyle{aasjournal}



\end{document}